\documentclass[epjc3]{svjour3} 

\usepackage[margin=25mm]{geometry}

\usepackage{amsmath,amssymb}

\usepackage[dvipdfmx]{graphicx} 
\usepackage{wrapfig}

\usepackage[xcolor,framemethod=tikz]{mdframed}
\usepackage{ulem}
\usepackage{cite}

\usepackage{url}

\usepackage{subfigure}

\newcommand\Eunq[1]{\langle #1 \rangle_{q}^{(2)}}
\newcommand\Eq[1]{\langle #1 \rangle^{\mathrm{(3)}}_{q}}
\newcommand{\ST}{S_{\mathrm{T}q}}
\newcommand{\hA}{\hat{A}}
\newcommand{\hH}{\hat{H}}
\newcommand{\qmrho}{\hat{\rho}}
\newcommand{\Tr}[1]{\mathrm{Tr}\left[#1\right]}
\newcommand{\betaPh}{\beta_{\mathrm{ph}}^{(3)}}
\newcommand{\tPh}{t_{\mathrm{ph}}^{(3)}}

\newcommand{\Utwo}{U^{(2)}}
\newcommand{\Uthree}{U^{(3)}}
\newcommand{\URthree}{U^{(3)}_{\mathrm{R}}}
\newcommand{\ER}{E_{\mathrm{R}}}
\newcommand{\STtwo}{S_{\mathrm{T}}^{(2)}}
\newcommand{\STthree}{S_{\mathrm{T}}^{(3)}}
\newcommand{\CvCEtwo}{C^{(2)}}
\newcommand{\CvCEthreePh}{C_{\mathrm{ph}}^{(3)}}
\newcommand{\gammatwo}{\gamma^{(2)}_N}
\newcommand{\gammathree}{\gamma^{(3)}_N}
\newcommand{\betatwo}{\beta^{(2)}}

\newcommand{\alphatwo}{\alpha^{(2)}_N}
\newcommand{\alphathree}{\alpha^{(3)}_N}
\newcommand{\numbertwo}{\overline{n}^{(2)}}
\newcommand{\numberthree}{\overline{n}^{(3)}}
\newcommand{\scTtwo}{t^{(2)}} 
 
\newcommand{\TPh}{T_{\mathrm{ph}}}

\newcommand{\Hzeta}[2]{\zeta_{\mathrm{H}}(#1,#2)}
\newcommand{\Bzeta}[3]{\zeta_{\mathrm{B}}(#1,#2|#3)}

\newcommand{\aeq}{a_{\mathrm{eq}}}
\newcommand{\beq}{b_{\mathrm{eq}}}
\def\<#1>{\langle #1 \rangle}

\newcommand\Econv[1]{\langle #1 \rangle^{(1)}}
\newcommand{\Uone}{U^{(1)}}
\newcommand{\STone}{S_{\mathrm{T}}^{(1)}}
\newcommand{\CvCEone}{C^{(1)}}

\newcommand{\betaone}{\beta^{(1)}}
\newcommand{\gammaone}{\gamma^{(1)}_N}
\newcommand{\numberone}{\overline{n}^{(1)}}
\newcommand{\scTone}{t^{(1)}} 

\title{Multiple quantum harmonic oscillators in the Tsallis statistics}
\author{Masamichi Ishihara\thanksref{e1,addr1}}
\thankstext{e1}{email: masamichi.ishihara.research@gmail.com}
\institute{Department of Economics, Faculty of Economics, Chiba Keizai University, Chiba, 263-0021, Japan \label{addr1}}

\begin{document}

\maketitle

\begin{abstract}
  We studied multiple quantum harmonic oscillators in the Tsallis statistics of entropic parameter
  $q$ in the cases that the distributions are power-like, separately applying
  the conventional expectation value, the unnormalized $q$-expectation value, and the normalized $q$-expectation value  (escort average).
  We obtained the expressions of the energy and the Tsallis entropy, using the Barnes zeta function.
  For the same oscillators,
  we obtained the expressions of the energy, the Tsallis entropy, the average level of the oscillators,
  and the heat capacity. 
  Numerically, we calculated the energy, the Tsallis entropy, and the heat capacity for various $N$ and $q$,
  using the expansion of the Barnes zeta function with the Hurwitz zeta function, 
  where $N$ is the number of independent oscillators. 
  The parameter $q$ is less than one in the Tsallis statistics with the conventional expectation value.
  The parameter $q$ is greater than one in both sets of the Tsallis statistics, each of which is defined with a different $q$-expectation value.
  These limitations of $q$ arise from the requirements that the distributions are power-like.
  It was shown from the requirements for the Barnes zeta function
  that $q$ is greater than $N/(N+1)$ for the conventional expectation value
  and that $q$ is less than $(N+1)/N$ for both of the $q$-expectation values.
  In the Tsallis statistics with the conventional expectation value,
  the energy, the Tsallis entropy, and the heat capacity decrease with $q$. 
  These quantities per oscillator increase with $N$.
  In the Tsallis statistics with the unnormalized $q$-expectation value,
  the energy, the Tsallis entropy, and the heat capacity increase with $q$ at low temperature,
  while decrease with $q$ at high temperature.
  These quantities per oscillator increase with $N$ at low temperature,
  while decrease with $N$ at high temperature.
  The heat capacity is the Schottky-type.
  The quantities are affected by the zero-point energy.
  In the Tsallis statistics with the normalized $q$-expectation value,
  the $N$ dependence of the energy per oscillator and that of the heat capacity per oscillator are quite weak, 
  and the $q$ dependence of the energy and that of the heat capacity are also weak,
  when the equilibrium temperature, which is often called the physical temperature, is adopted.
  The Tsallis entropy per oscillator decreases with $N$ and the Tsallis entropy decreases with $q$.
\end{abstract}

\section{Introduction}

Extension of the Boltzmann-Gibbs statistics has been attempted,
and one of extended statistics is the Tsallis statistics \cite{Tsallis:Book} which is widely used to describe various phenomena.
The Tsallis statistics is one parameter extension of the Boltzmann-Gibbs statistics,
and the introduced parameter $q$ is called entropic parameter.
This statistics is based on the Tsallis entropy and the expectation value.
The different expectation values are applied separately in the statistics:
the conventional expectation value, the unnormalized $q$-expectation value,
or the normalized $q$-expectation value  (escort average) is employed \cite{Tsallis:PhysicaA:1998}.
The conventional expectation value and 
the normalized $q$-expectation value have an appropriate property:
these expectation values of the unit operator $\hat{1}$ are one. 
The Tsallis statistics with the conventional expectation value is often called the Tsallis-1 statistics.
The parameter $q$ is less than one in the case that the distribution is power-like.
The Tsallis statistics with the unnormalized $q$-expectation value is often called the Tsallis-2 statistics, 
and the Tsallis statistics with the normalized $q$-expectation value is often called the Tsallis-3 statistics.
The parameter $q$ is greater than one in both of the statistics in the cases that the distributions are power-like.

The system of harmonic oscillators is basic, as the system of free particles is.
It is significant to calculate quantities for the systems in the Tsallis statistics.
The number of constituents is limited in the Tsallis statistics 
\cite{Abe-PLA:2001, Lenzi:PLA:2001, Ishihara:EPJB:2022, Ishihara:EPJB:2023},
while that is not limited in the Boltzmann-Gibbs statistics.
Such differences between the Tsallis statistics and the Boltzmann-Gibbs statistics may be clearly seen in simple systems.
Therefore, it is worth to study the system of independent constituents in the Tsallis statistics.

Heat capacity often appears in the study of the Tsallis statistics.
The Tsallis distribution appears in the system of the constant heat capacity \cite{Wada2003}.
The temperature fluctuation generates the $q$-exponential type distribution and is related to the heat capacity \cite{Wilk:EPJA40:2009}.
The condition between the entropic parameter and the heat capacity was also shown \cite{Ishihara:EPJP:03:2023}.
It is worth to study the heat capacity, because the heat capacity in the Tsallis statistics plays important roles.

It is not easy to calculate the quantities without approximations even for independent oscillators in the Tsallis statistics,
because the calculation for multiple oscillators cannot be decomposed into the calculation for a single oscillator.
This arises from the property of $q$-exponential function $\exp_q(x)$:
the $q$-exponential function satisfies $\exp_q(x+y +(1-q)xy) =  \exp_q(x) \exp_q(y)$ \cite{Tsallis:Book}.
Therefore, some approximations such as factorization approximation \cite{Buyukkilic:PLA:1995, Ubriaco:PRE:2000, Lenzi:PLA:2001},
high temperature approximation\cite{Ishihara:EPJB:2022}, or both are applied to proceed with calculations.
The validity of approximations was studied in the classical independent system of harmonic oscillators \cite{Lenzi:PLA:2001}.
The calculations without such approximations in quantum systems are required to avoid the effects of approximations.

The Barnes zeta function \cite{Ruijsenaars:2000,Kirsten:2010}
often appears in the calculations in the Tsallis statistics \cite{Ishihara:EPJB:2022,Ishihara:EPJB:2023,Oprisan}.
The value of a quantity can be estimated when the value of the Barnes zeta function is estimated.
The Barnes zeta function can be expanded with the Hurwitz zeta function \cite{Elizalde1989}, 
and the expansion is simplified in a certain case.
The representations of these zeta functions are useful.

The purpose of this paper is to study the system of oscillators in the Tsallis statistics.
The conventional expectation value, the unnormalized $q$-expectation value, and the normalized $q$-expectation value are employed separately. 
We calculate some quantities for independent oscillators of the total energy ${\displaystyle E=\sum_{j=1}^N (a_j n_j + b_j)}$,
where $N$ is the number of constituents.
The system of quantum harmonic oscillators has such energy.
Especially, we focus on the system of the oscillators with same frequency: $a_1=a_2=\cdots=a_N$. 
We represent the quantities without approximation, using the Barnes zeta function.
We calculate numerically the energy, the Tsallis entropy, and the heat capacity, 
using the expansion of the Barnes zeta function with the Hurwitz zeta function. 
The values of the parameters are chosen for each type of statistics to study the parameter dependences of the physical quantities,
because the values of the parameters are practically limited by numerical constraints.


We found the following facts.
In the Tsallis statistics of $q<1$ with the conventional expectation value, 
the parameter $q$ is greater than $N/(N+1)$.
The energy, the Tsallis entropy, and the heat capacity decrease with $q$.
These quantities per oscillator increase with $N$.
In the Tsallis statistics of $q>1$ with the unnormalized $q$-expectation value,
the parameter $q$ is less than $(N+1)/N$.
The energy, the Tsallis entropy, and the heat capacity increase with $q$ at low temperature,
while decrease with $q$ at high temperature.
These quantities per oscillator increase with $N$ at low temperature, while decrease with $N$ at high temperature.
The heat capacity is the Schottky-type: the heat capacity increases with the temperature, reaches the peak, and decreases after that.  
These quantities are affected by the zero-point energy.
In the Tsallis statistics of $q>1$ with the normalized $q$-expectation value,
the parameter $q$ is less than $(N+1)/N$.
The $N$ dependence of the energy per oscillator and that of the heat capacity per oscillator are quite weak, 
and the $q$ dependence of the energy and that of the heat capacity are also weak,
when the equilibrium temperature \cite{Imdiker:EPJC:2023,Ishihara:EPJP:2023:1,Ishihara:EPJP:2023:2},
which is often called the physical temperature
\cite{Abe-PLA:2001,Ishihara:EPJP:2023:1,Ishihara:EPJP:2023:2,Kalyana:2000,S.Abe:physicaA:2001,Aragao:2003,Ruthotto:2003,Toral:2003,Suyari:2006,Ishihara:phi4,Ishihara:free-field}, is adopted.
The equilibrium temperature dependence of the heat capacity in this statistics is similar to that in the Boltzmann-Gibbs statistics. 
In contrast, the Tsallis entropy per oscillator decreases with $N$ and the Tsallis entropy decreases with $q$.

This paper is organized as follows.
In Sec.~\ref{Sec:Basics}, we give a brief review of the Tsallis statistics.
The conventional expectation value, the unnormalized $q$-expectation value, and the normalized $q$-expectation value are introduced. 
We also give the expansion of the Barnes zeta function with the Hurwitz zeta function. 
In Sec.~\ref{Sec:Tsallis1}, Sec.~\ref{Sec:Tsallis2}, and Sec.~\ref{Sec:Tsallis3},
we deal with the energy, the Tsallis entropy, the average level of the oscillators, and the heat capacity
for multiple quantum harmonic oscillators in the Tsallis statistics.
The conventional expectation value is employed in Sec.~\ref{Sec:Tsallis1},
the unnormalized $q$-expectation value is employed in Sec.~\ref{Sec:Tsallis2},
and the normalized  $q$-expectation value is employed in Sec.~\ref{Sec:Tsallis3}. 
Last section is assigned for discussions and conclusions.
In \ref{sec:expansion},
we give the brief derivation of the expansion of the Barnes zeta function with the Hurwitz zeta function
according to the strategy given in the previous study. 
We also give a different method for deriving the expansion.

\section{Tsallis statistics and Barnes zeta function}
\label{Sec:Basics}
\subsection{Brief review of the Tsallis statistics} 
\label{Sec:formulation}
The Tsallis statistics is based on the Tsallis entropy and the expectation value \cite{Tsallis:Book, Tsallis:PhysicaA:1998}.
The Tsallis entropy of entropic parameter $q$ is defined by
\begin{align}
  \ST =  -\Tr{\qmrho^q \ln_q \qmrho} = - \Tr{\frac{\qmrho - \qmrho^q}{1-q}},
\end{align}
where $\qmrho$ is the density operator, 
$\mathrm{Tr}$ indicates the trace, and $\ln_q(x)$ is the $q$-logarithmic function. 
The $q$-logarithmic function \cite{Tsallis:Book,Suyari:2006} is defined by 
\begin{align}
  \ln_q(x) = \frac{x^{1-q} - 1}{1-q}, \qquad x>0. 
\end{align}  
The $q$-exponential function \cite{Tsallis:Book,Suyari:2006}, $\exp_q(x)$,  is defined by
\begin{align}
  \exp_q(x) =
  \left\{
  \begin{array}{ll}
    [ 1 + (1-q) x ]^{1/(1-q)} \qquad & 1 + (1-q) x > 0, \\
    0 & \mathrm{otherwise}. 
  \end{array}
  \right.
\end{align}

The first candidate for the expectation value of a quantity $\hA$ is given by
\begin{align}
\Econv{\hA} = \Tr{\qmrho \hA}. 
\label{Tsallis-1:expectation}
\end{align}
The Tsallis statistics with the conventional expectation value is often called the Tsallis-1 statistics.
The density operator $\qmrho^{(1)}$ in the Tsallis-1 statistics is given by applying the maximum entropy principle.
The following functional $I^{(1)}$ is extremized with the energy constraint $\Econv{\hH} = U$:
\begin{align}
I^{(1)} = \ST - \alpha^{(1)} (\Tr{\qmrho} - 1) - \beta^{(1)} (\Econv{\hH} - U),
\label{Tsallis1:functional}
\end{align}
where $\alpha^{(1)}$ and $\beta^{(1)}$ are the Lagrange multipliers.
The density operator $\qmrho^{(1)}$ is given by
\begin{subequations}
\begin{align}
  \qmrho^{(1)} &= \left[1 + (1-q) \ST +\left( \frac{1-q}{q} \right) \beta^{(1)} (\hat{H} - U) \right]^{1/(q-1)}.
  \label{Tsallis-1:density_op} 
\end{align}
The density operator can be rewritten when $1+(1-q)\ST$ is positive:
\begin{align}
  \qmrho^{(1)} &= \exp_{2-q}(-\ST) \exp_{2-q} \left(-\frac{\beta^{(1)}}{q (1+(1-q)\ST)} (\hat{H} - U) \right)
  = \frac{\exp_{2-q} \left(-\frac{\beta^{(1)}}{q (1+(1-q)\ST)} (\hat{H} - U) \right)}{\exp_{q}(\ST)} . 
\end{align}
\end{subequations}
We note that the Tsallis-1 statistics \cite{Parvan-1,Parvan-2} employed above
is different from that employed in the earlier work \cite{Tsallis:1988}.
The functional in the earlier work is different from Eq.~\eqref{Tsallis1:functional}.

The second candidate for the expectation value of a quantity $\hA$ is given by
\begin{align}
\Eunq{\hA} = \Tr{\qmrho^q \hA}. 
\label{Tsallis-2:expectation}
\end{align}
This is called the unnormalized $q$-expectation value. 
The Tsallis statistics with the unnormalized $q$-expectation value  is often called the Tsallis-2 statistics.
The density operator $\qmrho^{(2)}$ in the Tsallis-2 statistics is also given by applying the maximum entropy principle.
The following functional $I^{(2)}$ is extremized with the energy constraint $\Eunq{\hH} = U$:
\begin{align}
  I^{(2)} = \ST - \alpha^{(2)} (\Tr{\qmrho} - 1) - \beta^{(2)} (\Eunq{\hH} - U),
\end{align}
where $\alpha^{(2)}$ and $\beta^{(2)}$ are the Lagrange multipliers. 
The density operator $\qmrho^{(2)}$ is given by
\begin{subequations}
\begin{align}
  &\qmrho^{(2)} = \frac{1}{Z^{(2)}} \exp_q(-\beta^{(2)} \hH), \\
  &Z^{(2)} = \Tr{\exp_q(-\beta^{(2)} \hH)} . 
\end{align}
\label{Tsallis-2:density_op} 
\end{subequations}

The third candidate for the expectation value of a quantity $\hA$ is given by 
\begin{align}
  \Eq{\hA} =\frac{\Tr{\qmrho^q \hA}}{\Tr{\qmrho^q}}.  
\label{Tsallis-3:expectation}
\end{align}
This is called the normalized $q$-expectation value or the escort average. 
The Tsallis statistics with the normalized $q$-expectation value  is often called the Tsallis-3 statistics.
The density operator $\qmrho^{(3)}$ in the Tsallis-3 statistics is given in the same manner.
The following functional $I^{(3)}$ is extremized with the energy constraint $\Eq{\hH} = U$:
\begin{align}
I^{(3)} = \ST - \alpha^{(3)} (\Tr{\qmrho} - 1) - \beta^{(3)} (\Eq{\hH} - U),
\end{align}
where $\alpha^{(3)}$ and $\beta^{(3)}$ are the Lagrange multipliers. 
The density operator $\qmrho^{(3)}$ is given by
\begin{subequations}
\begin{align}
  &\qmrho^{(3)} = \frac{1}{Z^{(3)}} \exp_q\left(-\frac{\beta^{(3)}}{c_q} (\hH-U)\right), \\
  &Z^{(3)} = \Tr{\exp_q \left(-\frac{\beta^{(3)}}{c_q} (\hH-U)\right)}, \\
  &c_q \equiv \Tr{(\qmrho^{(3)})^q} .
\end{align}
\label{Tsallis-3:density_op} 
\end{subequations}
There is the following relation between $Z^{(3)}$ and $c_q$:
\begin{align}
c_q  = \left( Z^{(3)} \right)^{1-q} .
\end{align}
We introduce the notation $\betaPh$ which is the inverse of the equilibrium temperature.
The equilibrium temperature is often called the physical temperature.
The quantity $\betaPh$ is defined as
\begin{align}
\betaPh = \beta^{(3)}/c_q .
\end{align}

With these relations,
we calculate some quantities for multiple quantum harmonic oscillators in the Tsallis statistics,
applying separately the conventional expectation value, the unnormalized $q$-expectation value, and the normalized $q$-expectation value.

\subsection{The expansion of the Barnes zeta function with the Hurwitz zeta function}
\label{Subsec:Barnes_zeta}
In the present paper, we consider multiple quantum harmonic oscillators.
The total energy is given by
\begin{align}
E(\{n\}) = \sum_{j=1}^N (a_j n_j + b_j) , 
\label{eqn:En}
\end{align}
where
$a_j$ and $b_j$ are the coefficients related to the energy of the oscillator numbered $j$,
$N$ is the number of the oscillators, and $\{n\}$ represents the set of the quantum numbers, $n_1, \cdots, n_N$. 
We introduce $\aeq$ by $\aeq = a_1$ when the condition $a_1=a_2=\cdots=a_N$ is satisfied.
In the same way, we introduce $\beq$ by $\beq=b_1$ when the condition $b_1=b_2=\cdots=b_N$ is satisfied.

The Barnes zeta function appears in the Tsallis statistics when the energy, Eq.~\eqref{eqn:En}, is adopted.
The following equation appears:
\begin{align}
  \sum_{n_1=0, \cdots, n_N=0}^{\infty} \left[ 1+ \kappa  E(\{n\}) \right]^{-s}
  = \kappa^{-s} \sum_{n_1=0, \cdots, n_N=0}^{\infty}
  \left[ \frac{1 + \kappa \displaystyle \sum_{j=1}^N b_j}{\kappa} + \sum_{j=1}^N a_j n_j \right]^{-s} , 
  \label{example:calc}
\end{align}
where $s$ and $\kappa$ are the parameters. 
The right-hand side of Eq.~\eqref{example:calc} is represented with the Barnes zeta function.

We use the following expressions for the Hurwitz zeta function $\zeta_{\mathrm{H}}$ and the Barnes zeta function $\zeta_{\mathrm{B}}$. 
The Hurwitz zeta function $\Hzeta{s}{a}$ \cite{Shpot2016} is given by
\begin{align}
\Hzeta{s}{a} = \sum_{n=0}^{\infty} \frac{1}{(n+a)^s} .
\end{align}
The Barnes zeta function $\Bzeta{s}{a}{a_1,\cdots,a_N}$ is given by
\begin{align}
\Bzeta{s}{a}{a_1,\cdots,a_N} = \sum_{n_1=0,n_2=0,\cdots,n_N=0}^{\infty} \frac{1}{(a+ n_1 a_1 + \cdots + n_N a_N)^s}, 
\end{align}
where $a$, $a_1$, $\cdots$, $a_N$ are positive and $s$ is greater than $N$ \cite{Ruijsenaars:2000}.
The Barnes zeta function is a generalization of the Hurwitz zeta function.
We often write $\Bzeta{s}{a}{a_1,\cdots,a_N}$ as $\Bzeta{s}{a}{\vec{a}_N}$,
where $\vec{a}_N$ represents the set $(a_1, a_2, \cdots, a_N)$.
We also use the notation $\vec{1}_N$ to represent the set $(a_1=1, a_2=1, \cdots, a_N=1)$.
Therefore, the expression $\Bzeta{s}{a}{\vec{1}_N}$ indicates 
\begin{align}
\Bzeta{s}{a}{\vec{1}_N} = \sum_{n_1=0,n_2=0,\cdots,n_N=0}^{\infty} \frac{1}{(a+ n_1 + \cdots + n_N)^s}. 
\end{align}

As given in \ref{subsec:barnes:equal},
the Barnes zeta function can be represented with the Hurwitz zeta function.
The Barnes zeta function for $a_1=a_2=\cdots=a_N$ is given by
\begin{align}
  \left. \zeta_B(s, a| a_1, \cdots, a_N) \right|_{a_1=a_2=\cdots=a_N} 
  &= \frac{1}{(a_1)^s} \left[\zeta_H(s, d) + \sum_{p=1}^{\infty} \left( \begin{array}{c} N + p -2 \\ p \end{array} \right) \zeta_H(s, d + p )  \right],  
  \label{eqn:expansion-of-Barns-zeta}
\end{align}
where $d$ is defined by $d=a/a_1$ and $\left( \begin{array}{c} m \\ n \end{array}\right)$ is the number of combinations. 
Equation~\eqref{eqn:expansion-of-Barns-zeta} is used in numerical calculations to obtain the values of the Barnes zeta function.

\section{Multiple quantum harmonic oscillators in the Tsallis statistics with the conventional expectation value}
\label{Sec:Tsallis1}

In this section, 
we deal with the energy, the Tsallis entropy, the average level of the oscillators, and the heat capacity
for multiple quantum harmonic oscillators in the Tsallis-1 statistics:
we employ the Tsallis entropy and the conventional expectation value. 
The probability in the Tsallis-1 statistics employed in this study is invariant to energy shift.
We calculate the energy, the Tsallis entropy, and the heat capacity numerically.

\subsection{Physical quantities for multiple quantum harmonic oscillators in the Tsallis statistics with the conventional expectation value} 
\label{subsec:Tsallis1:Theory}

We attempt to obtain the probability $p^{(1)}$, the energy $\Uone$, and the Tsallis entropy $\STone$
by using Eqs.~\eqref{Tsallis-1:density_op} and \eqref{Tsallis-1:expectation}.
We obtain the expressions of the physical quantities for the energy, Eq.~\eqref{eqn:En}.
The Barnes zeta function appears as shown in Subsec.~\ref{Subsec:Barnes_zeta}:
\begin{subequations}
\begin{align}
  & p^{(1)} = \left[ 1+(1-q) \STone + \left( \frac{1-q}{q} \right) \betaone (E(\{n\}) - \Uone) \right]^{1/(q-1)}, \\
  &\Uone_R = B^{1/(q-1)} \left\{ \Bzeta{q/(1-q)}{R/B - \Uone_R}{\vec{a}_N} + \left(\Uone_R - \frac{R}{B}\right) \Bzeta{1/(1-q)}{R/B - \Uone_R}{\vec{a}_N} \right\},\\
  &R = B^{\frac{q}{1-q}} \Bzeta{q/(1-q)}{R/B-\Uone_R}{\vec{a}_N},
\end{align}
\end{subequations}
where $\Uone_R$, $R$, and $B$ are defined by 
\begin{subequations}
\begin{align}
  &\Uone_R  = \Uone - \sum_{j=1}^N b_j , \label{eqn:U:UR}\\
  &R = 1 + (1-q) \STone , \label{eqn:R:S}\\
  &B = \frac{1-q}{q} \betaone .
\end{align}
\end{subequations}
The energy $\Uone$ is given with Eq.~\eqref{eqn:U:UR} and the Tsallis entropy $\STone$ is given with Eq.~\eqref{eqn:R:S}.
The parameter $q$ is less than one in the case that the distribution is power-like.
The parameter $q$ is greater than $N/(N+1)$ from the requirement $q/(1-q) > N$ for the Barnes zeta function. 
These limitations indicate $N/(N+1) < q < 1$. 

We focus on the case where $a_1, \cdots, a_N$ are all equal to $\aeq$.
It is worth to mention that no condition for $b_j$ is imposed.
We have the probability $p^{(1)}$, the scaled energy $u_R^{(1)} = \Uone_R/\aeq$, the Tsallis entropy $\STone$,
and the average level $\numberone$ which is defined by
\begin{align}
\numberone = \frac{1}{N} \sum_{n_1=0, \cdots, n_N=0}^{\infty} \left( \sum_{j=1}^N n_j \right)  p^{(1)} .
\end{align}
The heat capacity $\CvCEone$ is defined by 
\begin{align}
  \CvCEone  = \frac{\partial \Uone}{\partial T^{(1)}},
\end{align}
where $T^{(1)}$ is given by $1/\betaone$.

The quantities, $p^{(1)}$, $u_R^{(1)}$, and $R$, are given by
\begin{subequations}
  \begin{align}
  & p^{(1)} = Q^{\frac{1}{1-q}} \Big( QR - u_R^{(1)} + \sum_{j=1}^N n_j \Big)^{\frac{1}{(q-1)}}, \\
  & u_R^{(1)} =  Q R + Q^{\frac{1}{1-q}} \  (u_R^{(1)} - QR)\  \Bzeta{1/(1-q)}{QR-u_R^{(1)}}{\vec{1}_N}, \label{eqn:ur}\\  
  & R = Q^{\frac{q}{1-q}}  \Bzeta{q/(1-q)}{QR-u_R^{(1)}}{\vec{1}_N}, \label{eqn:R}
\end{align}
\end{subequations}
where the scaled temperature $t^{(1)}$ and the quantity $Q$ are given by
\begin{subequations}
  \begin{align}
    &\scTone =\frac{1}{\aeq \betaone} ,\\
    & Q \equiv Q(\scTone) = \frac{q\scTone}{1-q} =\frac{1}{\aeq B} .
\end{align}
\end{subequations}
The scaled energy $u_R^{(1)}$ and the entropy $\STone$ are obtained numerically from Eqs.~\eqref{eqn:ur} and \eqref{eqn:R}.
In order to solve the equations numerically, it is better to introduce the quantity $\gammaone$ defined by $\gammaone=QR-u_R^{(1)}$.
The quantity $\gammaone$ should be positive from the requirement for the Barnes zeta function $\Bzeta{s}{\gammaone}{\vec{1}_N}$.
We have the equations:
\begin{subequations}
  \begin{align}
  & 0 =  1 - Q^{\frac{1}{1-q}} \  \Bzeta{1/(1-q)}{\gammaone}{\vec{1}_N}, \label{eqn:x:ur}\\  
  & R = Q^{\frac{q}{1-q}}  \Bzeta{q/(1-q)}{\gammaone}{\vec{1}_N}. \label{eqn:x:R}
\end{align}
\end{subequations}
We can numerically calculate the scaled energy $u_R^{(1)}$ and the entropy $\STone$ 
from these equations by obtaining $\gammaone$.

The average level is given by 
\begin{align}
  \numberone &= N^{-1} \left[
    Q R + Q^{\frac{1}{1-q}} \  (u_R^{(1)} - QR)\  \Bzeta{1/(1-q)}{QR-u_R^{(1)}}{\vec{1}_N}
    \right]. \label{Tsallis1:n}
\end{align}
It is easily found from Eqs.~\eqref{eqn:ur} and \eqref{Tsallis1:n} that $\numberone$ equals $u_R^{(1)}/N$.
We can show the relation between $\Uone$ and $\numberone$ directly.
With the definitions of $\Uone$ and $\numberone$, we have
\begin{align}
  \Uone = \sum_{\{n\}} \sum_{j=1}^N (\aeq n_j + b_j) p^{(1)}(\{n\})
  = \aeq \sum_{\{n\}} \Big(\sum_{j=1}^N  n_j\Big) p^{(1)}(\{n\}) + \sum_{j=1}^N  b_j
  = N \aeq \numberone + \sum_{j=1}^N  b_j . 
\end{align}
This equation indicate $\numberone = u_R^{(1)}/N$.
The simple relation between $\Uone$ and $\numberone$ is obtained in the case that $b_1$, $\cdots$, $b_N$ are all equal to $\beq$:
\begin{align}
  \Uone  = N (\aeq \numberone + \beq).
\end{align}
The natural relation between the energy and the average level holds in the Tsallis-1 statistics.

The heat capacity is given by 
\begin{align}
  \CvCEone
  &= \frac{q}{(1-q)^2} \left(\frac{q \scTone}{1-q} \right)^{\frac{q}{1-q}} \Bzeta{q/(1-q)}{\gammaone}{\vec{1}_N}
  - \frac{1}{\scTone (1-q) \left(\frac{q\scTone}{1-q} \right)^{1/(1-q)} \Bzeta{(2-q)/(1-q)}{\gammaone}{\vec{1}_N}}.
\end{align}
The heat capacity is obtained with the solution $\gammaone$ of Eq.~\eqref{eqn:x:ur}.

\subsection{Numerical results in the Tsallis statistics with the conventional expectation value}

In the Tsallis statistics with the conventional expectation value,
we treat the case where $a_1$, $a_2$, $\cdots$, $a_N$ are all equal to $\aeq$. 
We calculate the scaled energy $\Uone_R/\aeq$, the Tsallis entropy $\STone$, and the heat capacity $\CvCEone$
as functions of the scaled temperature $t^{(1)}$ numerically.
We choose the value of $q$ to satisfy the inequality $N/(N+1) < q < 1$, because the Barnes zeta function requires $q/(1-q) > N$.

\begin{figure}
  \centering
  \subfigure[The scaled energies]
            {
            \includegraphics[width=0.45\textwidth]{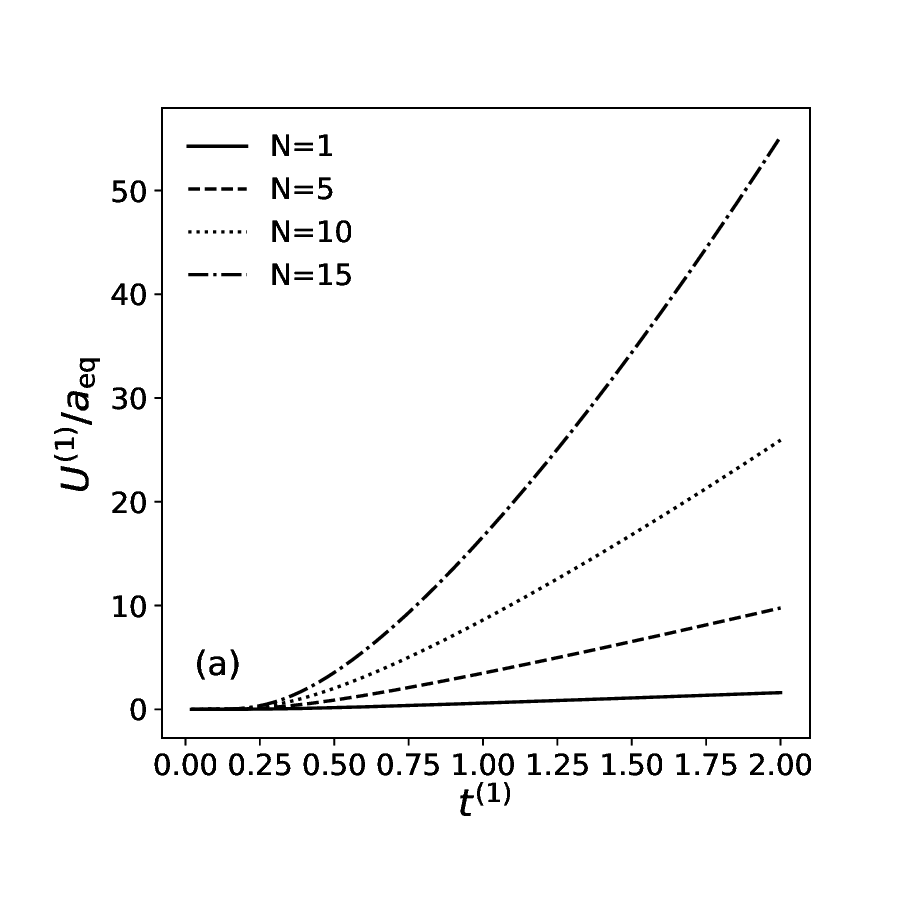}
            \label{fig:1:U1:a}
            }
            \hfill
  \subfigure[The scaled energies divided by $N$]
  {
            \includegraphics[width=0.45\textwidth]{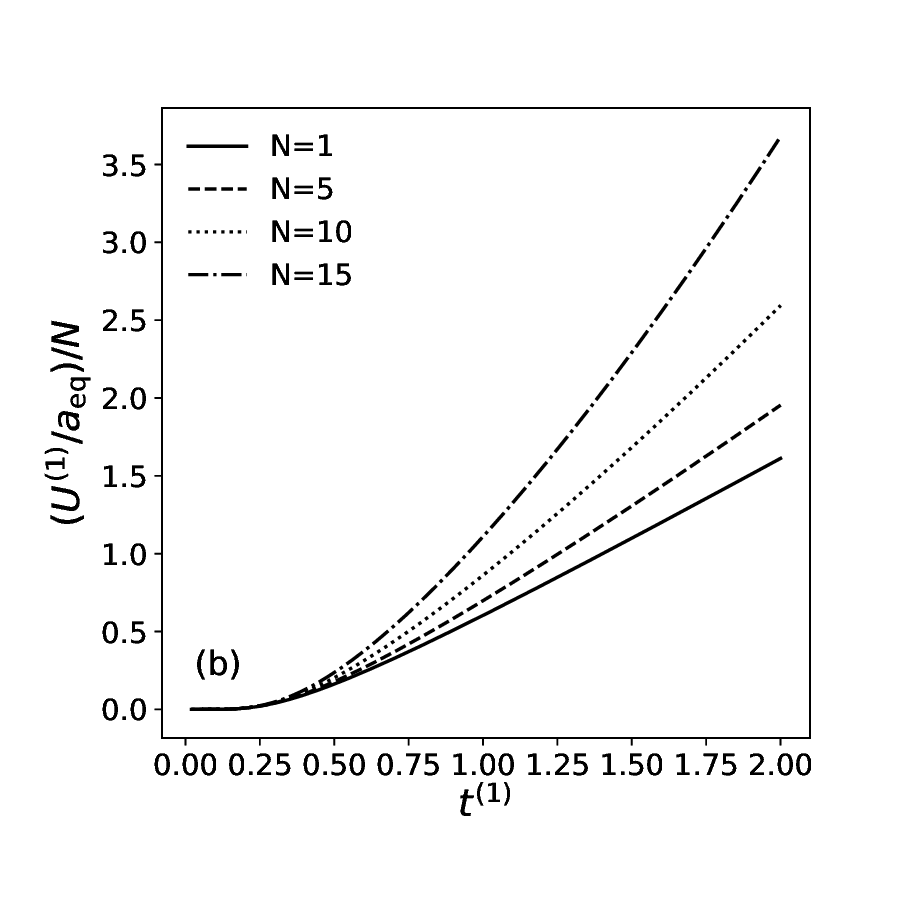}
            \label{fig:1:U1:b}
            }
  \caption{The scaled energies $\Uone/\aeq$ and the scaled energies divided by $N$, $(\Uone/\aeq)/N$, 
    as functions of the scaled temperature $t^{(1)}$ at $q=0.98$ for $N=1$, $5$, $10$, and $15$.}
  \label{fig:1:U1:Ndep}
\end{figure}

\begin{figure}[tbp]
    \centering
    \includegraphics[width=0.47\textwidth]{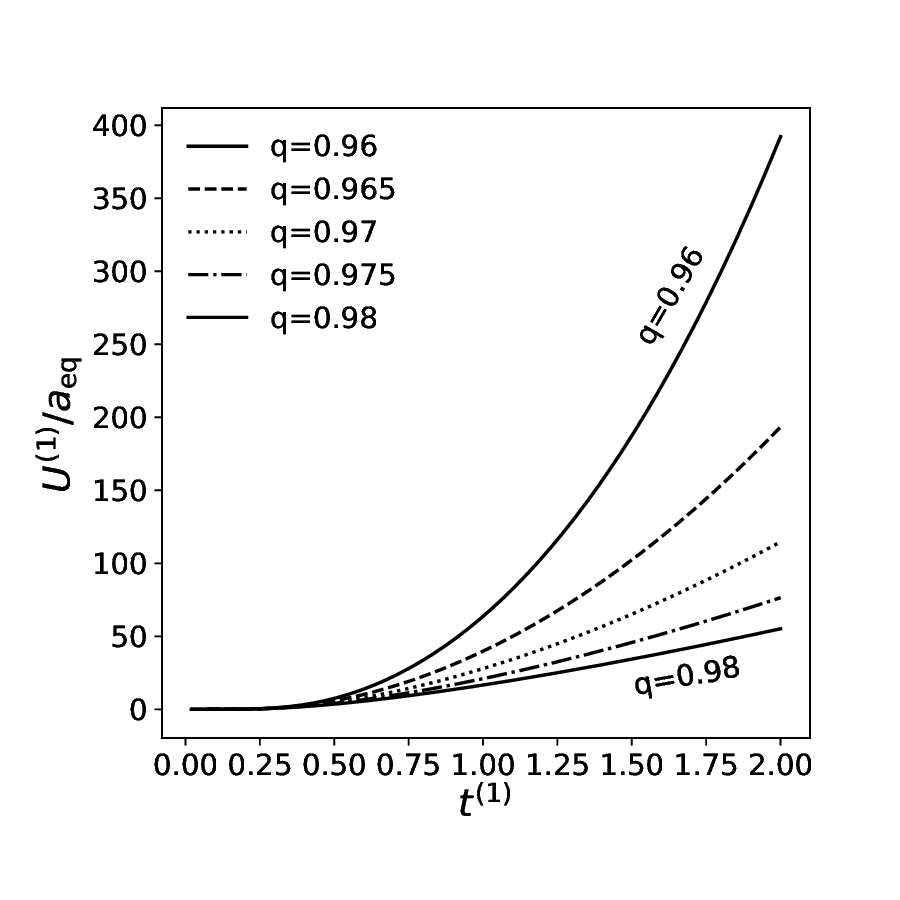}
    \caption{The scaled energies $\Uone/\aeq$ as functions of the scaled temperature $t^{(1)}$ at $N=15$
      for $q=0.96$, $0.965$, $0.97$, $0.975$, and $0.98$.}
  \label{fig:1:U1:qdep}
\end{figure}

First, we calculate the scaled energy numerically. 
Figure~\ref{fig:1:U1:a} shows the scaled energies $\Uone_R/\aeq$ as functions of $t^{(1)}$ at $q=0.98$ for $N=1, 5, 10$, and $15$. 
Figure~\ref{fig:1:U1:b} shows the scaled energies divided by $N$, $(\Uone/\aeq)/N$,
as functions of $t^{(1)}$ at $q=0.98$ for $N=1, 5, 10$, and $15$.
The energy increases with $t^{(1)}$, and the energy per oscillator increases with $N$.
Figure~\ref{fig:1:U1:qdep} shows the scaled energies $\Uone/\aeq$ at $N=15$ for $q=0.96$, $0.965$, $0.97$, $0.975$ and $0.98$.
The energy decreases with $q$.

\begin{figure}
  \centering
  \subfigure[The Tsallis entropies]
            {
              \includegraphics[width=0.45\textwidth]{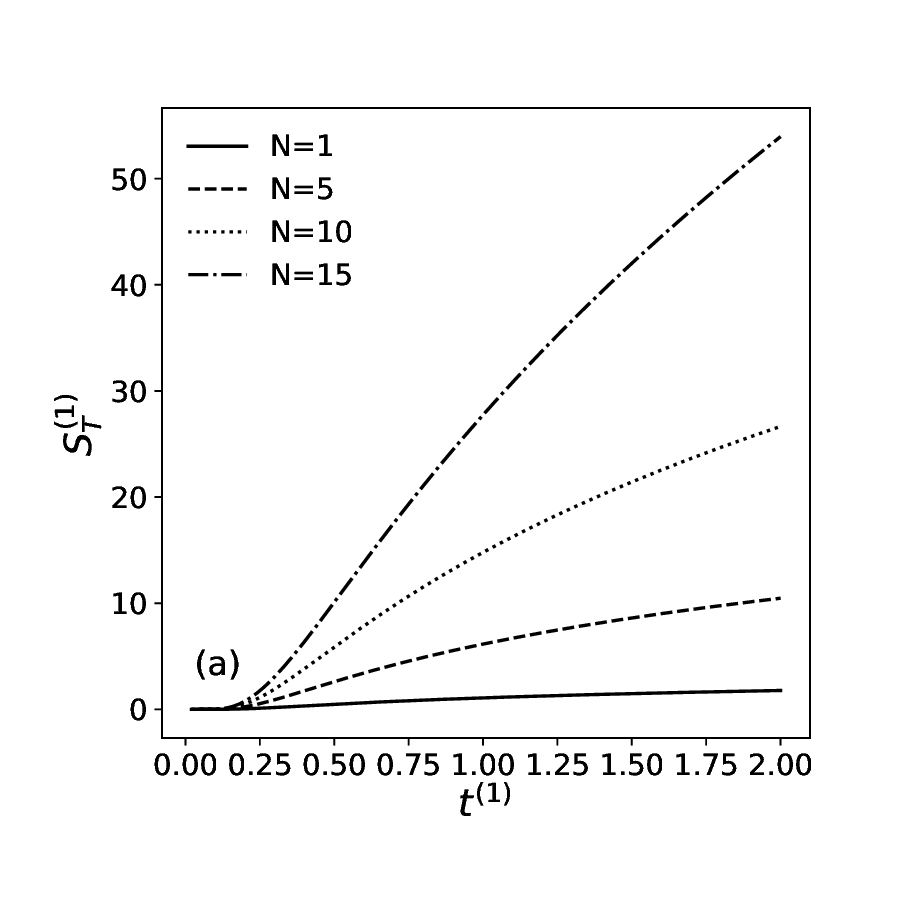} 
            \label{fig:StOne:Ndep:normal}
            }
            \hfill
  \subfigure[The Tsallis entropies divided by $N$]
            {
              \includegraphics[width=0.45\textwidth]{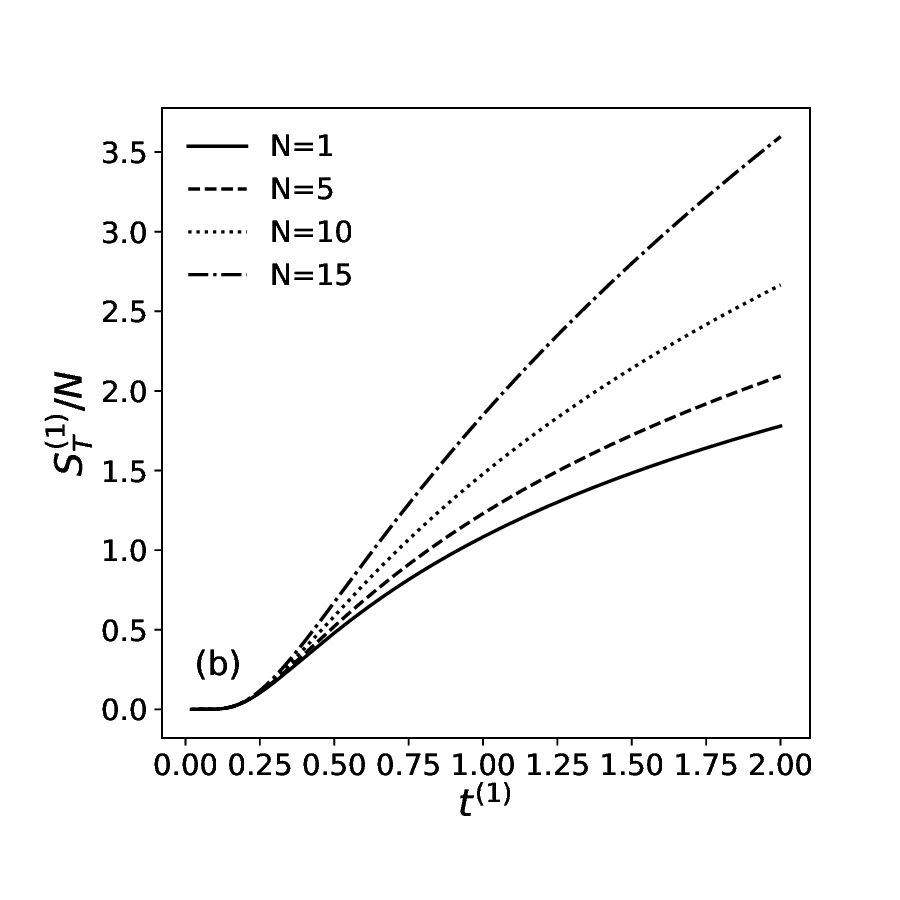}
            \label{fig:StOne:rescaled}
            }
            \caption{The Tsallis entropies $\STone$ and the Tsallis entropies divided by $N$, $\STone/N$, 
              as functions of the scaled temperature $t^{(1)}$ at $q=0.98$ for $N=1$, $5$, $10$, and $15$.}
            \label{fig:StOne:Ndep}
\end{figure}

\begin{figure}[tbp]
    \centering
    \includegraphics[width=0.47\textwidth]{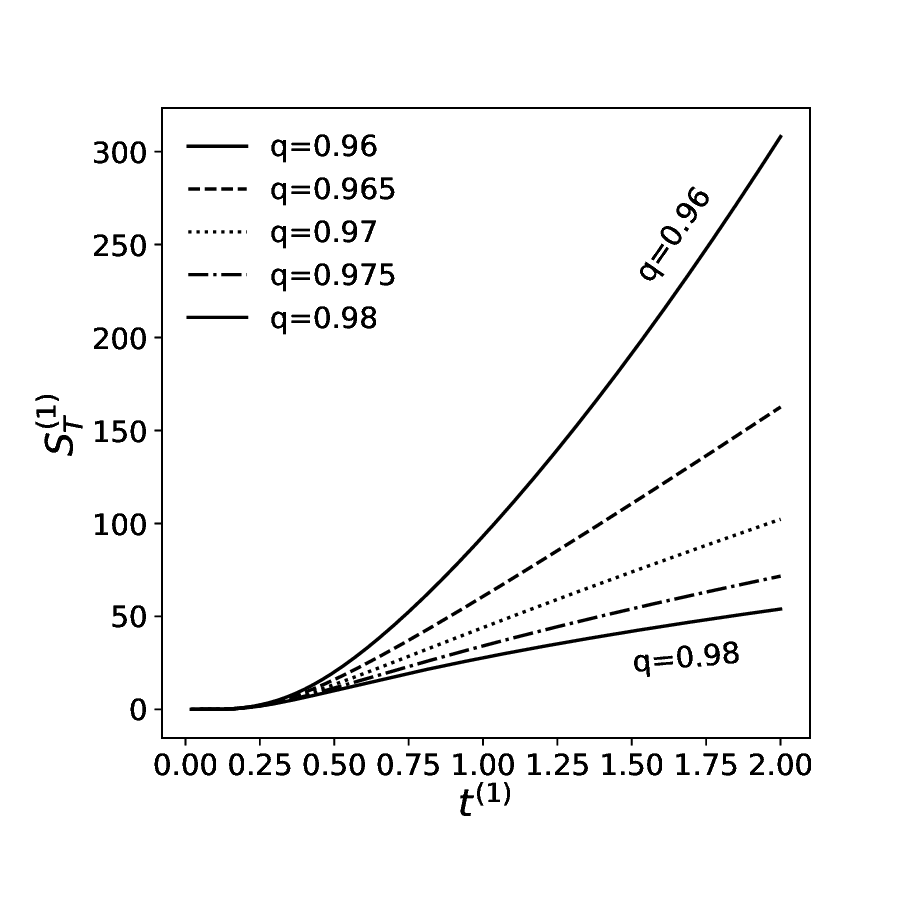}
    \caption{The Tsallis entropies $\STone$ as functions of the scaled temperature $t^{(1)}$ at $N=15$ for $q=0.96$, $0.965$, $0.97$, $0.975$, and $0.98$.}
    \label{fig:StOne:qdep}
\end{figure}

Next, we calculate the Tsallis entropy numerically. 
Figure~\ref{fig:StOne:Ndep:normal} shows the Tsallis entropies $\STone$
as functions of $t^{(1)}$ at $q=0.98$ for $N=1$, $5$, $10$, and $15$.
Figure~\ref{fig:StOne:rescaled} shows the Tsallis entropies divided by $N$, $\STone/N$,
as functions of $t^{(1)}$ at $q=0.98$ for $N=1$, $5$, $10$, and $15$.
The Tsallis entropy per oscillator increases with $N$.
Figure~\ref{fig:StOne:qdep} shows the Tsallis entropies $\STone$ as functions of $t^{(1)}$
at $N=15$ for $q=0.96$, $0.965$, $0.97$, $0.975$, and $0,98$.
The Tsallis entropy decreases with $q$.


\begin{figure}
  \centering
  \subfigure[The heat capacities] 
            {
              \includegraphics[width=0.45\textwidth]{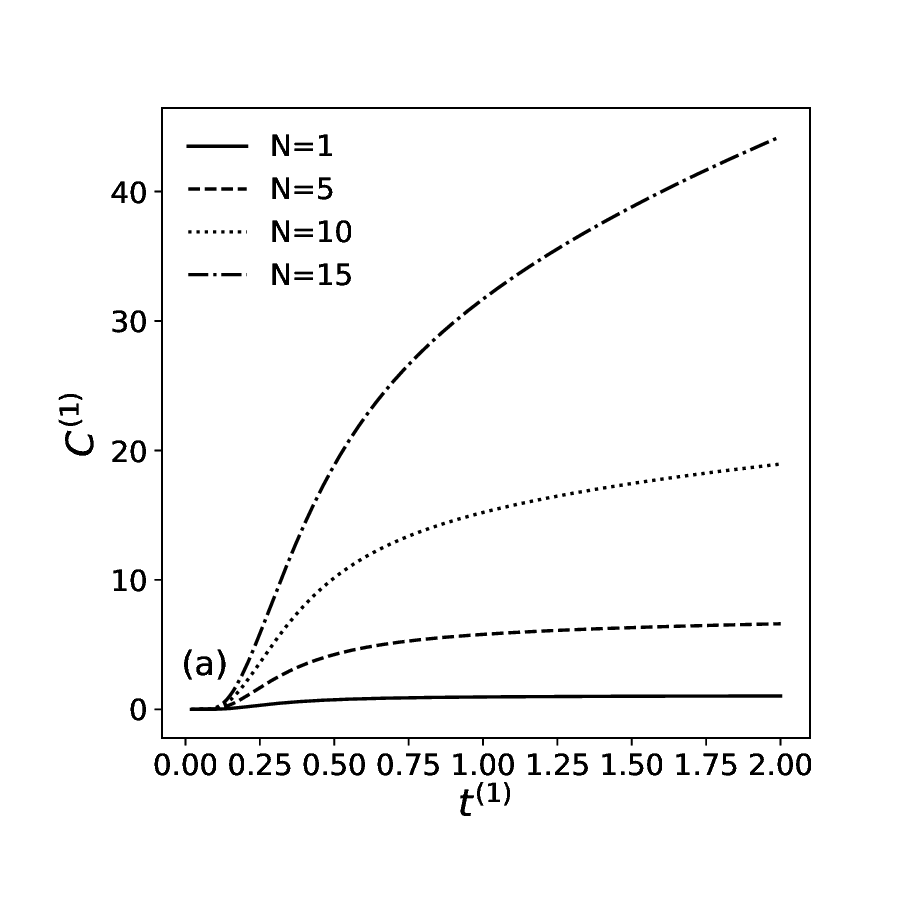}
              \label{fig:1:Cv:Ndep}
            }
            \hfill
            \subfigure[The heat capacities divided by $N$]
            {
              \includegraphics[width=0.45\textwidth]{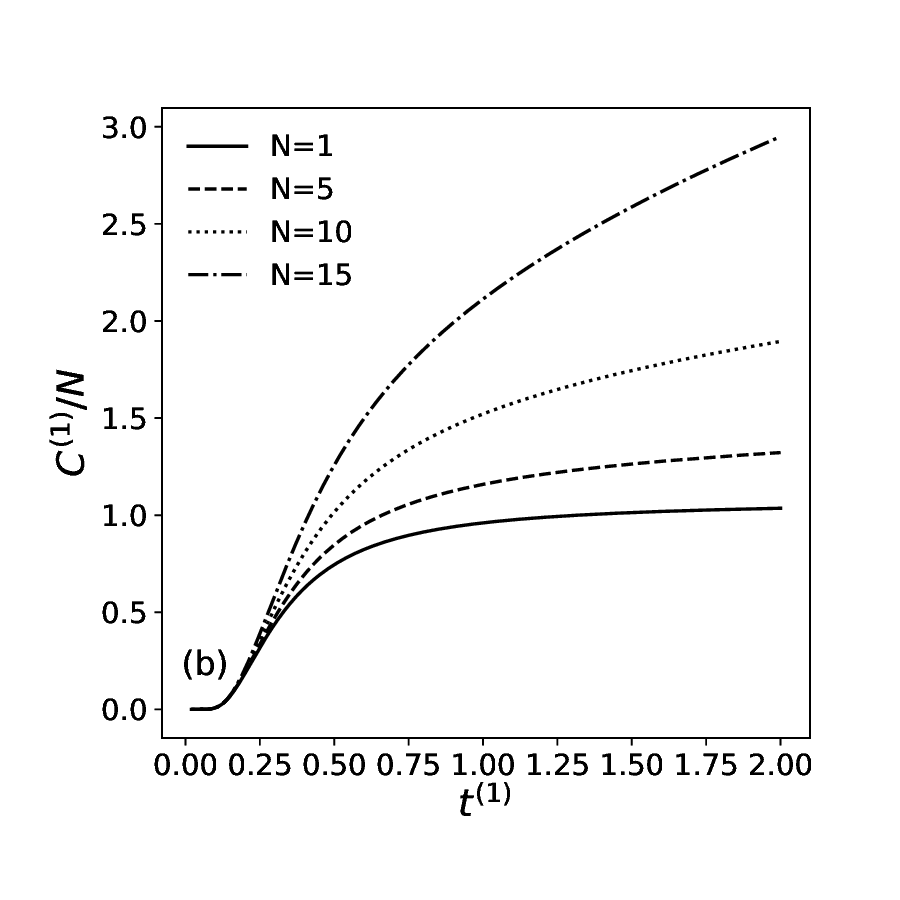}
              \label{fig:1:Cv:Ndep:b}
            }
            \caption{
              The heat capacities $\CvCEone$ and the heat capacities divided by $N$, $\CvCEone/N$,
              as functions of the scaled temperature $t^{(1)}$ at $q=0.98$ for $N=1$, $5$, $10$, and $15$.
            }
\end{figure}

\begin{figure}[tbp]
    \centering
    \includegraphics[width=0.47\textwidth]{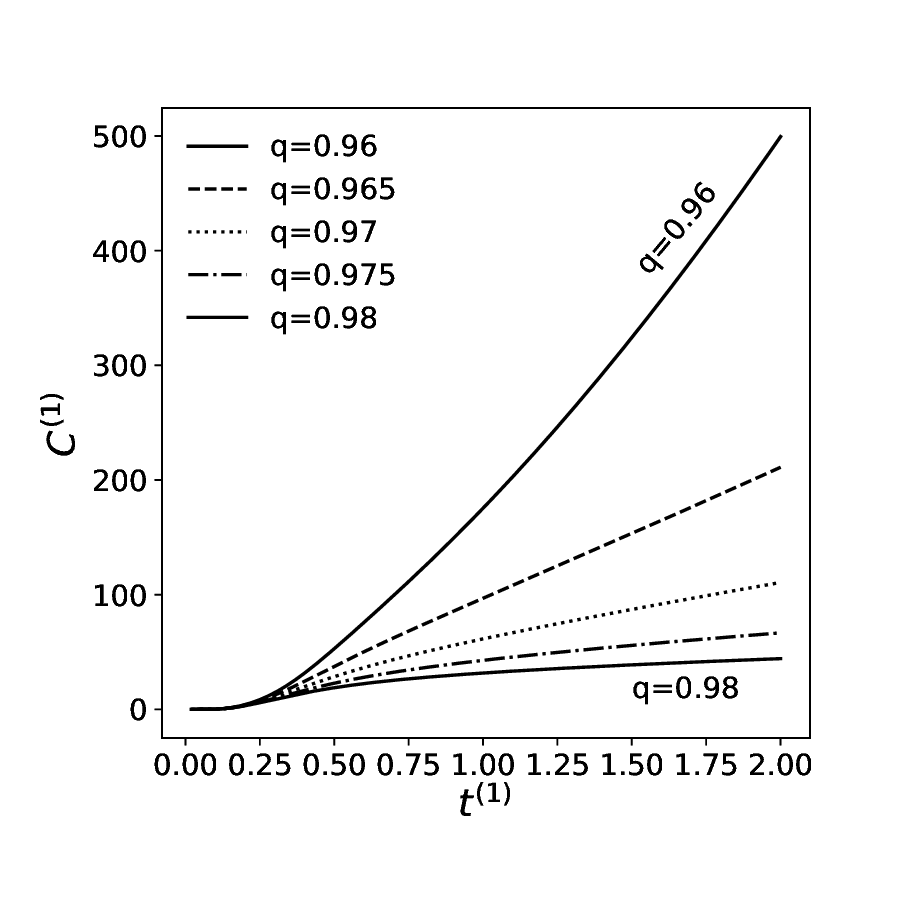}
    \caption{The heat capacities $\CvCEone$ as functions of the scaled temperature $t^{(1)}$ at $N=15$ for $q=0.96$, $0.965$, $0.97$, $0.975$, and $0.98$.}
  \label{fig:1:Cv:qdep}
\end{figure}

Finally, we calculate the heat capacity numerically. 
Figure~\ref{fig:1:Cv:Ndep} shows the heat capacities $\CvCEone$
as functions of $t^{(1)}$ at $q=0.98$ for $N=1$, $5$, $10$, and $15$.
The heat capacity increases with $t^{(1)}$.
Figure~\ref{fig:1:Cv:Ndep:b} shows the heat capacities divided by $N$, $\CvCEone/N$, 
as functions of $t^{(1)}$ at $q=0.98$ for $N=1$, $5$, $10$, and $15$.
The heat capacity per oscillator, $\CvCEone/N$, increases with $N$.
Figure~\ref{fig:1:Cv:qdep} shows the heat capacities $\CvCEone$
as functions of $t^{(1)}$ at $N=15$ for $q=0.96$, $0.965$, $0.97$, $0.975$, and $0.98$.
The heat capacity decreases with $q$. 

\section{Multiple quantum harmonic oscillators in the Tsallis statistics with the unnormalized $q$-expectation value }
\label{Sec:Tsallis2}

In this section,
we deal with the energy, the Tsallis entropy, the average level of the oscillators, and the heat capacity 
for multiple quantum harmonic oscillators in the Tsallis-2 statistics:
we employ the Tsallis entropy and the unnormalized $q$-expectation value.
We calculate the energy, the Tsallis entropy, and the heat capacity numerically.

\subsection{Physical quantities for multiple quantum harmonic oscillators in the Tsallis statistics with the unnormalized $q$-expectation value} 
\label{subsec:Tsallis2:Theory}


We attempt to obtain the probability $p^{(2)}$, the energy $\Utwo$, and the Tsallis entropy $\STtwo$
by using Eqs.~\eqref{Tsallis-2:density_op} and \eqref{Tsallis-2:expectation}. 
We obtain the expressions of the physical quantities for the energy, Eq.~\eqref{eqn:En}:
\begin{subequations}
\begin{align}
  & p^{(2)} = \frac{\left[\alphatwo + \displaystyle\sum_{j=1}^N a_j n_j\right]^{1/(1-q)}}{\Bzeta{1/(q-1)}{\alphatwo}{\vec{a}_N}}, \\
  &\Utwo =\frac{1}{\left[\Bzeta{1/(q-1)}{\alphatwo}{\vec{a}_N}\right]^{q-1}} - 
  \frac{\Bzeta{q/(q-1)}{\alphatwo}{\vec{a}_N}}{\left[\Bzeta{1/(q-1)}{\alphatwo}{\vec{a}_N}\right]^{q}}, \\
  &\STtwo
= \frac{1}{q-1} \left\{1-\frac{\Bzeta{q/(q-1)}{\alphatwo}{\vec{a}_N}}{\left[\Bzeta{1/(q-1)}{\alphatwo}{\vec{a}_N}\right]^{q}}\right\},
\end{align}
\end{subequations}
where $\alphatwo$ is given by
\begin{align}
\alphatwo = \frac{1+(q-1)\betatwo \displaystyle\sum_{j=1}^N b_j }{(q-1)\betatwo} .
\end{align}
The parameter $q$ is greater than one in the case that the distribution is power-like.
The parameter $q$ is less than $(N+1)/N$ from the requirement $1/(q-1) > N$ for the Barnes zeta function. 
These limitations indicate $1 < q < (N+1)/N$.

We focus on the case where $a_1, \cdots, a_N$ are all equal to $\aeq$ and $b_1, \cdots, b_N$ are all equal to $\beq$.
We have the probability $p^{(2)}$, the scaled energy $\Utwo/\aeq$, the Tsallis entropy $\STtwo$,
and the average level $\numbertwo$ which is defined by
\begin{align}
\numbertwo = \frac{1}{N} \sum_{n_1=0, \cdots, n_N=0}^{\infty} \left( \sum_{j=1}^N n_j \right) \left( p^{(2)} \right)^q .
\end{align}
The heat capacity $\CvCEtwo$ is defined by 
\begin{align}
  \CvCEtwo  = \frac{\partial \Utwo}{\partial T^{(2)}},
\end{align}
where $T^{(2)}$ equals $1/\betatwo$. These quantities are given by
\begin{subequations}
\begin{align}
  &p^{(2)} = \frac{\left[\gammatwo + \displaystyle\sum_{j=1}^N n_j\right]^{1/(1-q)}}{\Bzeta{1/(q-1)}{\gammatwo}{\vec{1}_N}},\\
  &\frac{\Utwo}{\aeq}
  = \frac{1}{\left[\Bzeta{1/(q-1)}{\gammatwo}{\vec{1}_N}\right]^{q-1}}
  - \frac{\scTtwo}{(q-1)} \frac{\Bzeta{q/(q-1)}{\gammatwo}{\vec{1}_N}}{\left[\Bzeta{1/(q-1)}{\gammatwo}{\vec{1}_N}\right]^q},  \label{Tsallis2:u}\\
  &\STtwo = \frac{1}{q-1} \left\{1-\frac{\Bzeta{q/(q-1)}{\gammatwo}{\vec{1}_N}}{\left[\Bzeta{1/(q-1)}{\gammatwo}{\vec{1}_N}\right]^{q}}\right\}, \label{Tsallis2:S}\\
  &\numbertwo = \frac{1}{N}
  \left\{
  \frac{1}{\left[\Bzeta{1/(q-1)}{\gammatwo}{\vec{1}_N}\right]^{q-1}}
  - \gammatwo \frac{\Bzeta{q/(q-1)}{\gammatwo}{\vec{1}_N}}{\left[\Bzeta{1/(q-1)}{\gammatwo}{\vec{1}_N}\right]^q}
  \right\}, \label{Tsallis2:n}\\
  &\CvCEtwo
  = \frac{q  \scTtwo}{(q-1)^3}  \frac{1}{\left[ \Bzeta{1/(q-1)}{\gammatwo}{\vec{1}_N} \right]^{q+1} }\\  
  & \qquad\qquad \times \left\{
  \Bzeta{1/(q-1)}{\gammatwo}{\vec{1}_N}  \Bzeta{(2q-1)/(q-1)}{\gammatwo}{\vec{1}_N} - \left[\Bzeta{q/(q-1)}{\gammatwo}{\vec{1}_N} \right]^2 
  \right\}, 
\end{align}
\end{subequations}
where $t^{(2)}$ and $\gammatwo$ are given by
\begin{subequations}
\begin{align}
  &\scTtwo =\frac{1}{\aeq \betatwo} ,\\
  &\gammatwo = \frac{t^{(2)}}{(q-1)} + N \left( \frac{\beq}{\aeq} \right) \label{eqn:Tsallis2:gammatwo}.
\end{align}
\end{subequations}
The quantity $\beq/\aeq$ should be greater than or equal to zero for $\scTtwo>0$ in order for $\gammatwo$ to be positive.

We find the following relation from Eqs.~\eqref{Tsallis2:u}, \eqref{Tsallis2:S}, and \eqref{Tsallis2:n}:
\begin{align}
  \Utwo  = N (\aeq \numbertwo + \beq) + (1-q) N \beq \STtwo.
  \label{relation:Utwho:n:STtwo}
\end{align}
This equation is directly derived with the definitions of $\Utwo$, $\STtwo$, and $\numbertwo$:
\begin{align}
\Utwo &= \aeq \sum_{\{n\}} \Big( \sum_{j=1}^N n_j \Big) \big(p^{(2)}\big)^q + \beq N \sum_{\{n\}} \big(p^{(2)}\big)^q 
= N \aeq \numbertwo + \beq N \big(1+(1-q) \STtwo \big) \nonumber \\
&= N (\aeq \numbertwo + \beq) + (1-q) N \beq \STtwo. 
\end{align}  
Equation~\eqref{relation:Utwho:n:STtwo} is derived again.
In the Boltzmann-Gibbs limit ($q \rightarrow 1$), we have the natural relation:
\begin{align}
\Utwo  = N (\aeq \numbertwo + \beq) .
\end{align}  

\subsection{Numerical results in the Tsallis statistics with the unnormalized $q$-expectation value}
In the Tsallis statistics with the unnormalized $q$-expectation value,
we treat the case where $a_1$, $a_2$, $\cdots$, $a_N$ are all equal to $\aeq$
and $b_1$, $b_2$, $\cdots$, $b_N$ are all equal to $\beq$.
We calculate the scaled energy $\Utwo/\aeq$, the Tsallis entropy $\STtwo$, and the heat capacity $\CvCEtwo$
as functions of the scaled temperature $t^{(2)}$ numerically.
We choose the value of $q$ to satisfy the inequality $1 < q < (N+1)/N$,
because the Barnes zeta function requires $1/(q-1) > N$.

\begin{figure}
  \centering
  \subfigure[The scaled energies]
  {
            \includegraphics[width=0.45\textwidth]{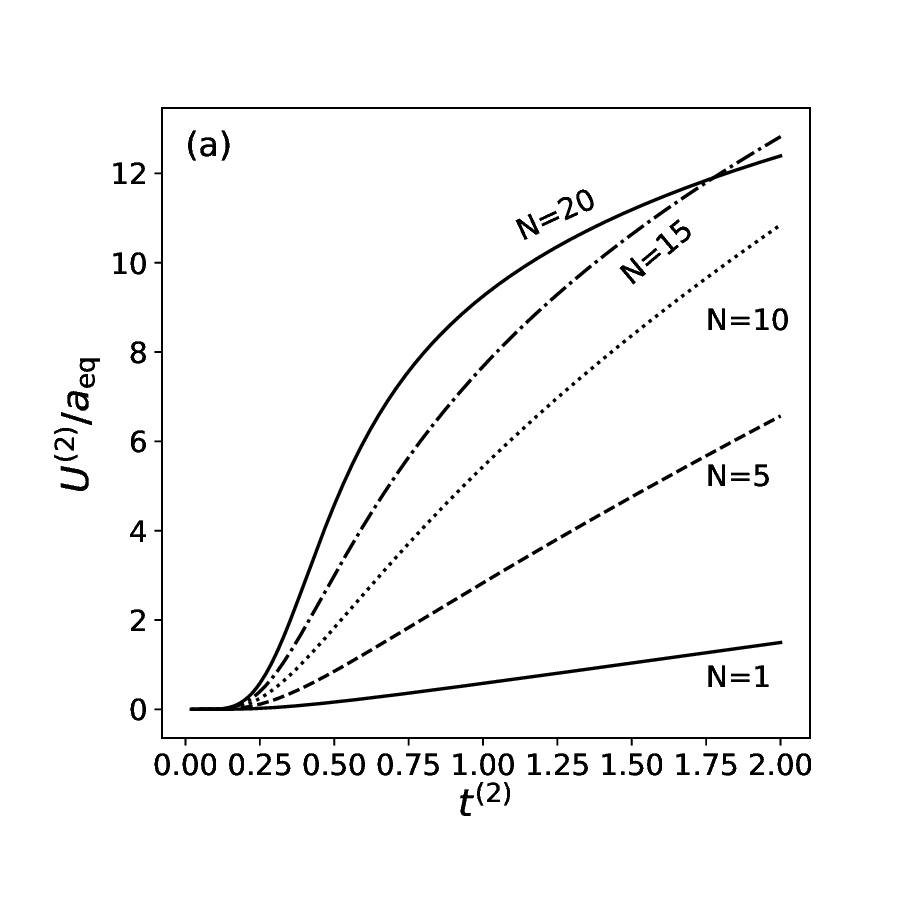}
            \label{fig:2:U2:a}
            }
            \hfill
  \subfigure[The scaled energies divided by $N$]
  {
            \includegraphics[width=0.45\textwidth]{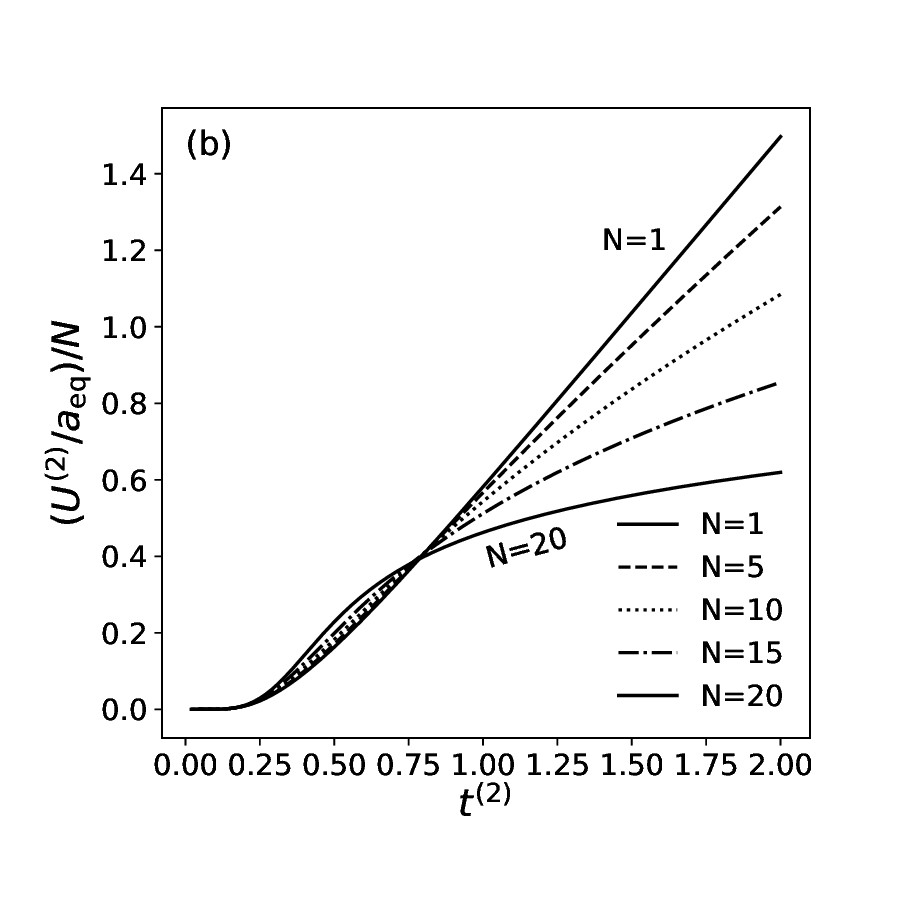}
            \label{fig:2:U2:b}
            }

  \subfigure[The scaled energies divided by $N$ in the narrow range]
  {
            \includegraphics[width=0.45\textwidth]{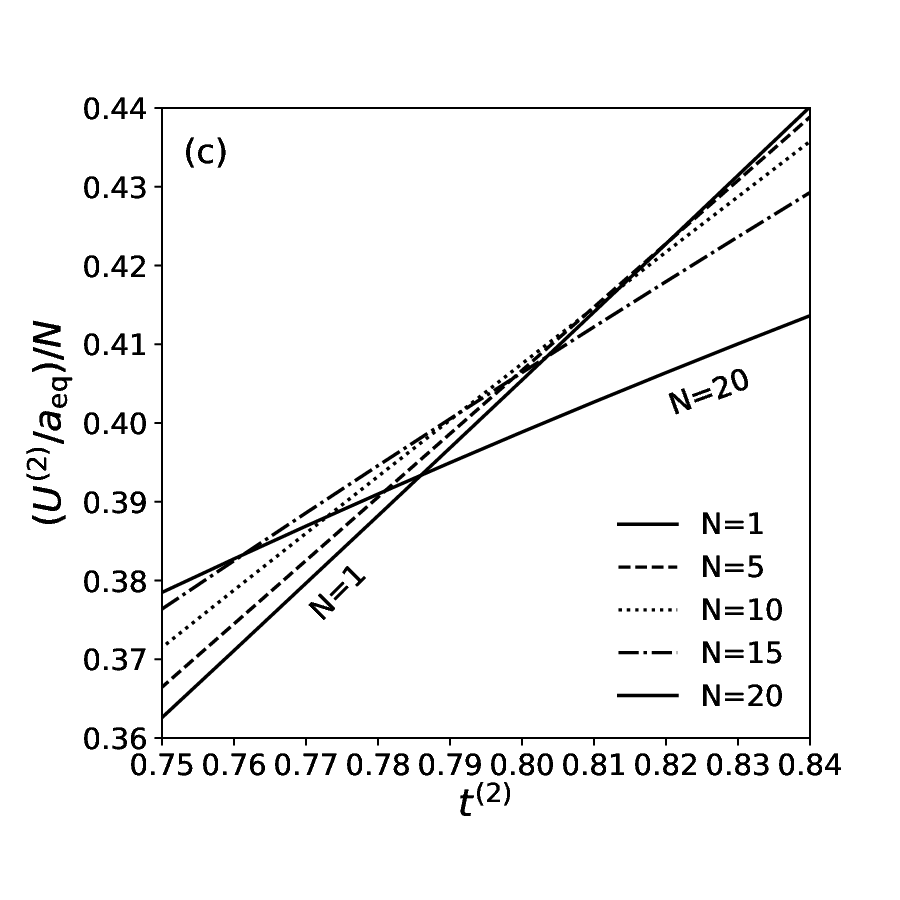}
            \label{fig:2:U2:c}
            }

  \caption{The scaled energies $\Utwo/\aeq$, the scaled energies divided by $N$, $(\Utwo/\aeq)/N$,
    and the scaled energies divided by $N$ in the narrow range, 
    as functions of the scaled temperature $t^{(2)}$ at $q=1.04$ and $\beq/\aeq=0.0$
    for $N=1$, $5$, $10$, $15$, and $20$.}
  \label{fig:2:U2:Ndep}
\end{figure}

\begin{figure}[tbp]
  \begin{minipage}[b]{0.48\columnwidth}
    \centering
    \includegraphics[width=0.94\textwidth]{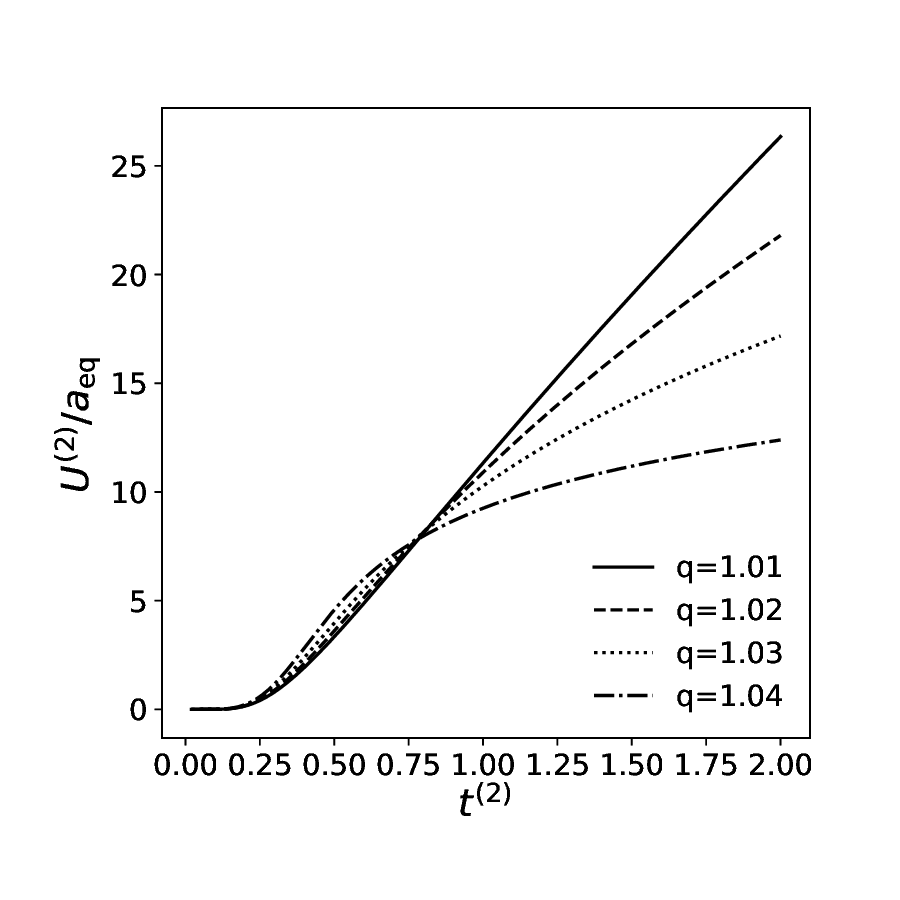} 
    \caption{The scaled energies $\Utwo/\aeq$ as functions of the scaled temperature $t^{(2)}$ at $N=20$ and $\beq/\aeq=0.0$
      for $q=1.01$, $1.02$, $1.03$, and $1.04$.}
  \label{fig:2:U2:qdep}
  \end{minipage}
  \hspace{0.04\columnwidth} 
  \begin{minipage}[b]{0.48\columnwidth}
    \centering
    \includegraphics[width=0.94\textwidth]{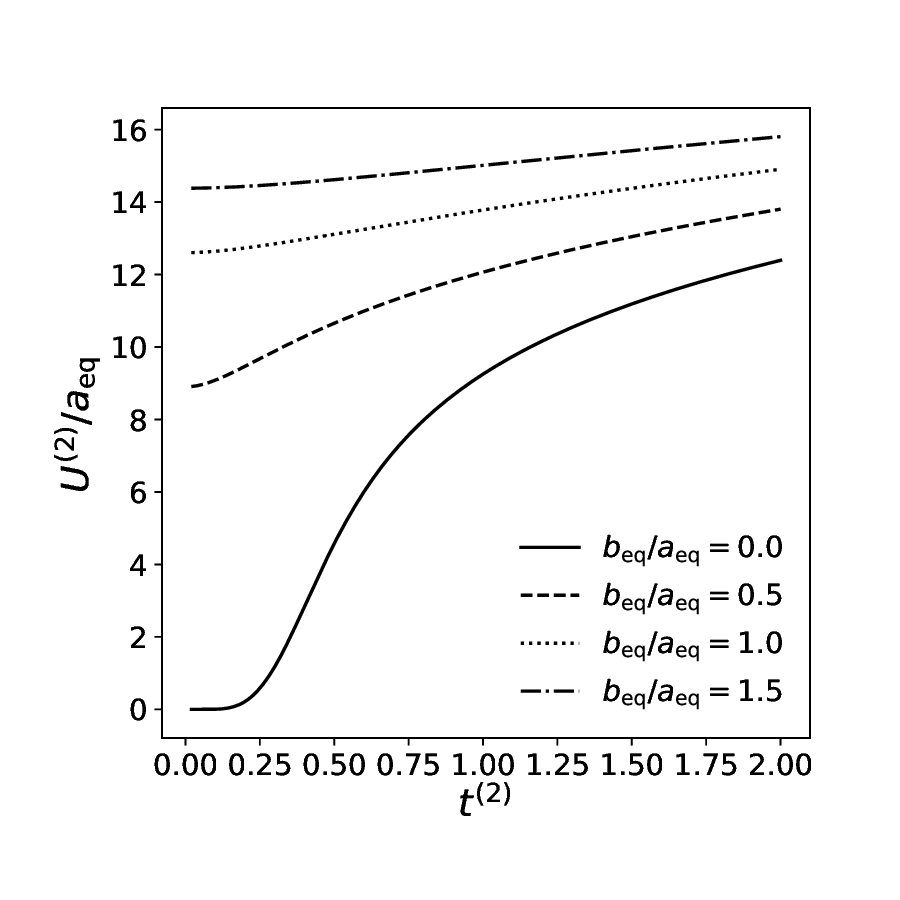} 
    \caption{The scaled energies $\Utwo/\aeq$ as functions of the scaled temperature $t^{(2)}$ at $q=1.04$ and $N=20$
      for $\beq/\aeq=0.0$, $0.5$, $1.0$, and $1.5$.}
  \label{fig:2:U2:bdep}
  \end{minipage}
\end{figure}

First, we calculate the scaled energy numerically. 
Figure~\ref{fig:2:U2:a} shows the scaled energies $\Utwo/\aeq$ as functions of $t^{(2)}$
at $q=1.04$ and $\beq/\aeq=0.0$ for $N=1, 5, 10, 15,$ and $20$. 
Figures~\ref{fig:2:U2:b} and \ref{fig:2:U2:c} show the scaled energies divided by $N$, $(\Utwo/\aeq)/N$,
as functions of $t^{(2)}$ at $q=1.04$ and $\beq/\aeq=0.0$ for $N=1, 5, 10, 15,$ and $20$.
The scaled energy increases with $\scTtwo$. 
The scaled energy per oscillator decreases with $N$ at high scaled temperature,
while the scaled energy per oscillator increases with $N$ at low scaled temperature. 
It seems that a fixed point exists in Fig.~\ref{fig:2:U2:b}.   
However, this point is not a fixed point as shown in Fig.~\ref{fig:2:U2:c}.
Figure~\ref{fig:2:U2:qdep} shows the scaled energy $\Utwo/\aeq$ at $N=20$ and $\beq/\aeq=0.0$ for $q=1.01$, $1.02$, $1.03$, and $1.04$.
The energy decreases with $q$ at high scaled temperature,
while the energy increases with $q$ at low scaled temperature.
A fixed point does not exist in Fig.~\ref{fig:2:U2:qdep},
even though it seems that the lines intersect at the same point.  
Figure~\ref{fig:2:U2:bdep} shows the scaled energy $\Utwo/\aeq$ at $q=1.04$ and $N=20$ for $\beq/\aeq=0.0$, $0.5$, $1.0$, and $1.5$.
The energies decrease and approach non-negative values as $\scTtwo$ decreases, depending on the value of $\beq/\aeq$. 

\begin{figure}
  \centering
  \subfigure[The Tsallis entropies]
            {
              \includegraphics[width=0.45\textwidth]{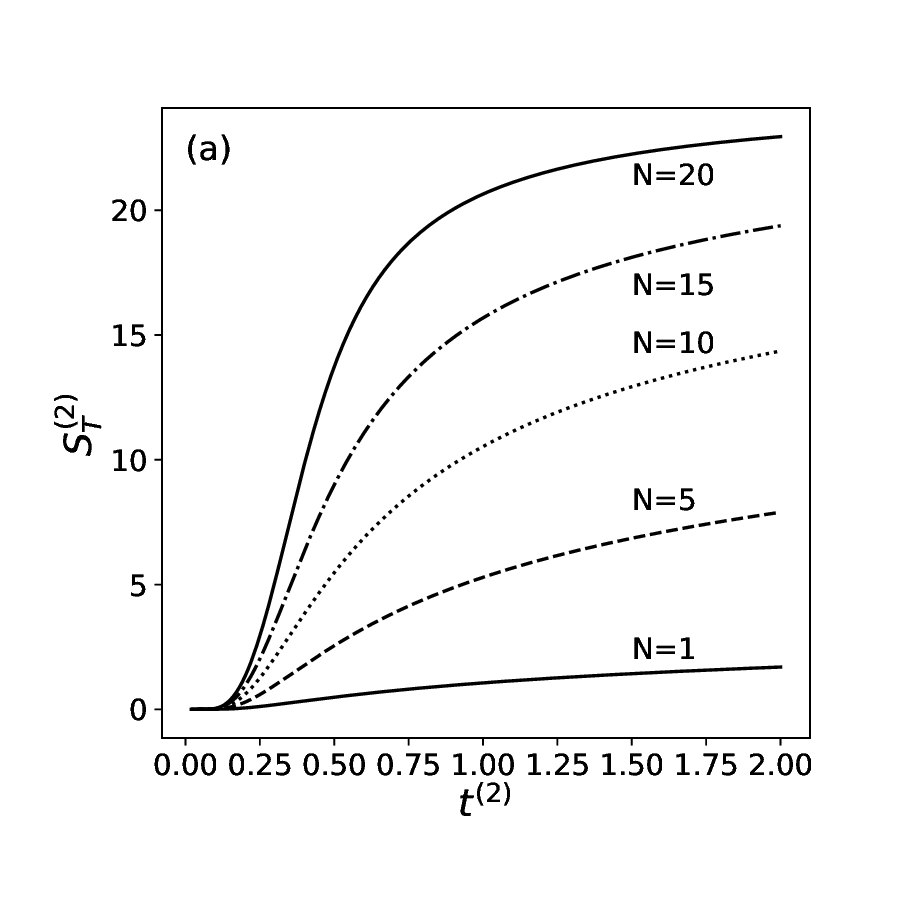}
            \label{fig:StTwo:Ndep:normal}
            }
            \hfill
  \subfigure[The Tsallis entropies divided by $N$]
            {
              \includegraphics[width=0.45\textwidth]{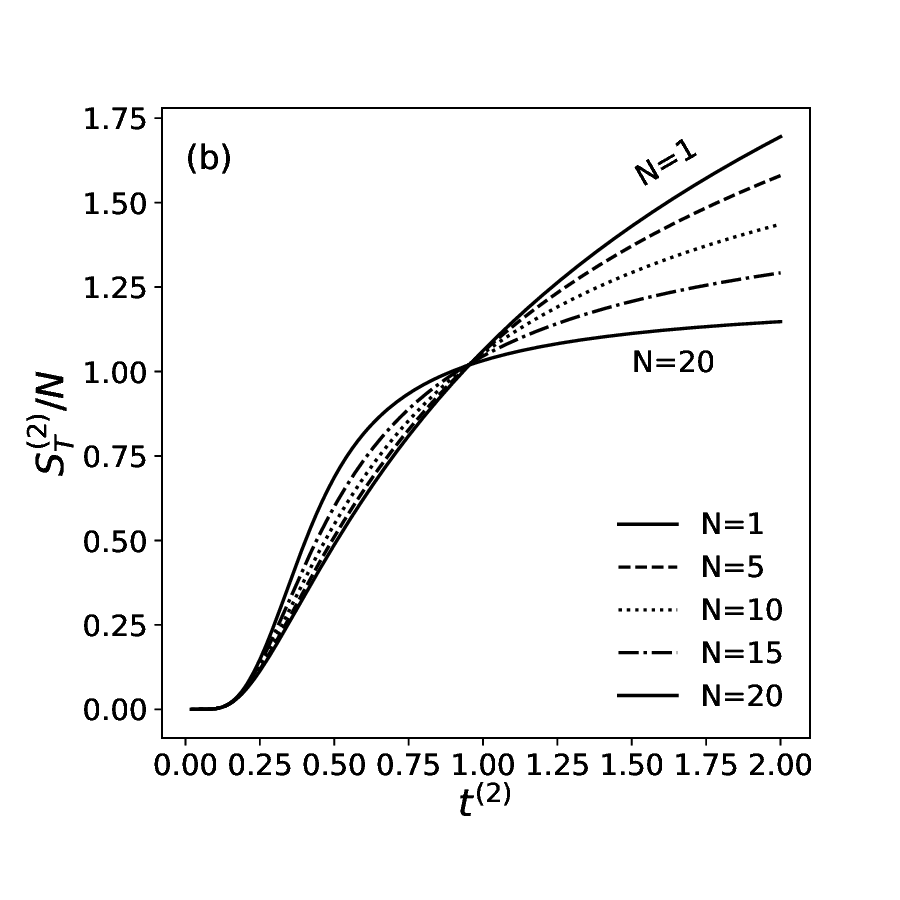}
            \label{fig:StTwo:rescaled}
            }
            \caption{The Tsallis entropies $\STtwo$ and the Tsallis entropies divided by $N$, $\STtwo/N$, 
              as functions of the scaled temperature $t^{(2)}$ at $q=1.04$ and $\beq/\aeq=0.0$
              for $N=1$, $5$, $10$, $15$, and $20$.}
            \label{fig:StTwo:Ndep}
\end{figure}

\begin{figure}[tbp]
  \begin{minipage}[b]{0.48\columnwidth}
    \centering
    \includegraphics[width=0.94\textwidth]{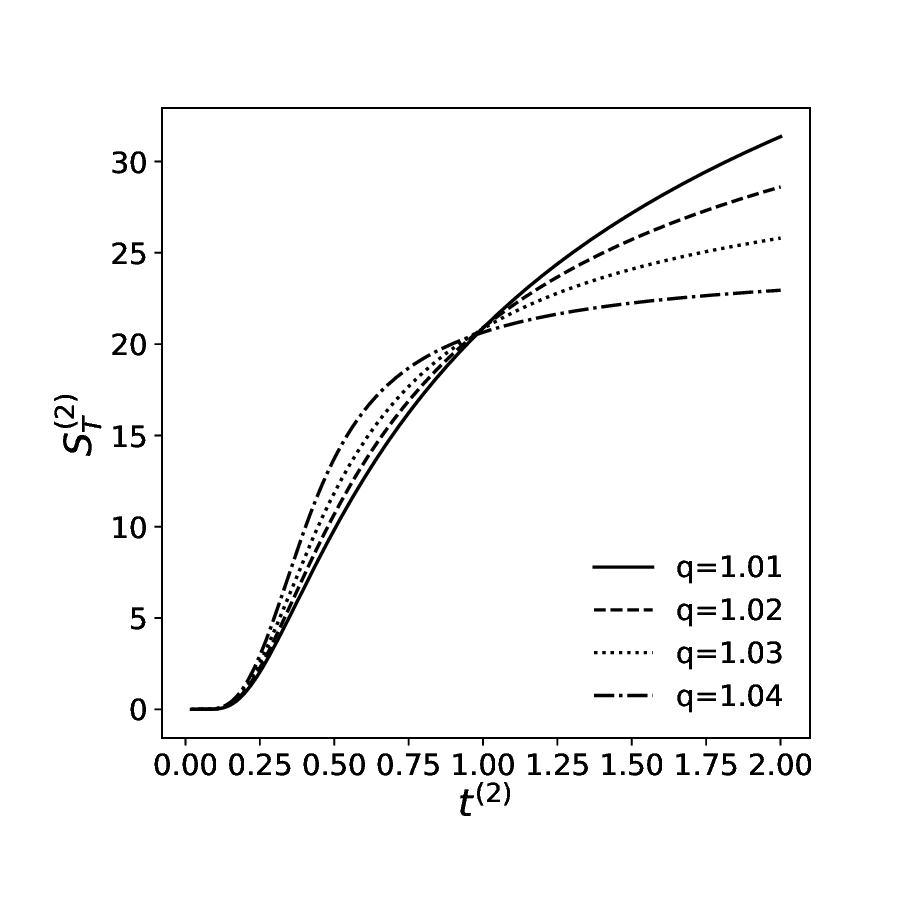}
    \caption{The Tsallis entropies $\STtwo$ as functions of the scaled temperature $t^{(2)}$ at $N=20$ and $\beq/\aeq=0.0$
      for $q=1.01$, $1.02$, $1.03$, and $1.04$.}
    \label{fig:StTwo:qdep}
  \end{minipage}
  \hspace{0.04\columnwidth} %
  \begin{minipage}[b]{0.48\columnwidth}
    \centering
    \includegraphics[width=0.94\textwidth]{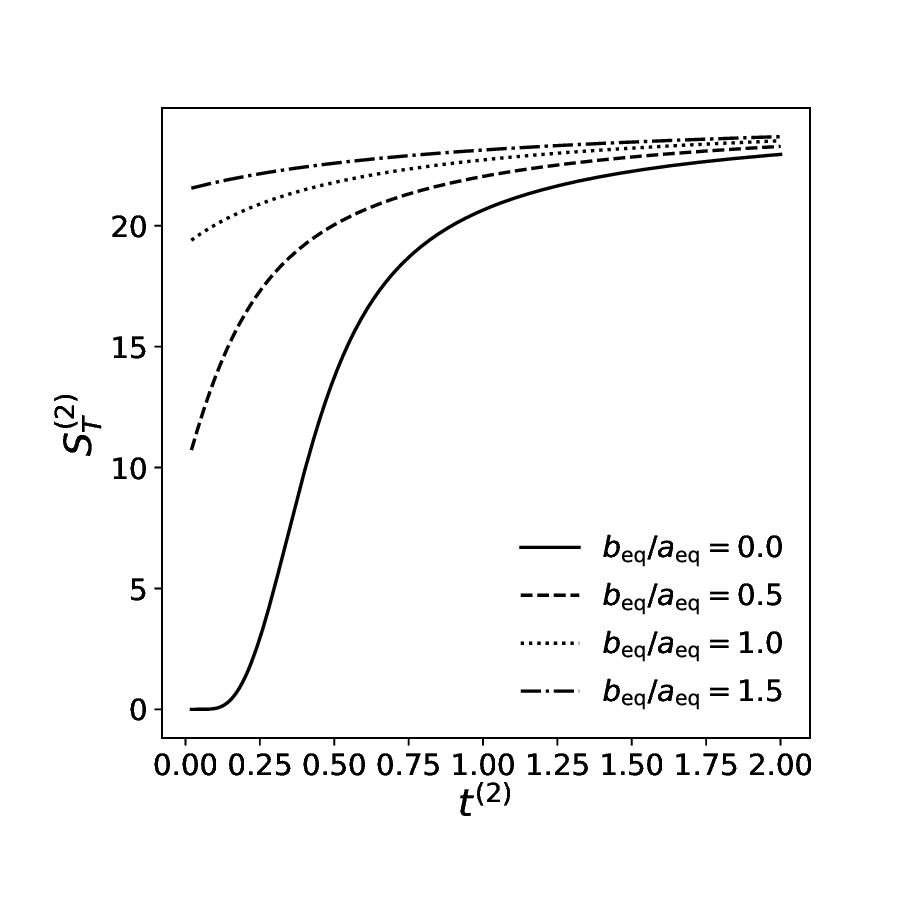}
    \caption{The Tsallis entropies $\STtwo$ as functions of the scaled temperature $t^{(2)}$ at $q=1.04$ and $N=20$
      for $\beq/\aeq=0.0$, $0.5$, $1.0$, and $1.5$.
    }
  \label{fig:StTwo:bdep}
  \end{minipage}
\end{figure}

Next, we calculate the Tsallis entropy numerically. 
Figure~\ref{fig:StTwo:Ndep:normal} shows the Tsallis entropies $\STtwo$
as functions of $t^{(2)}$ at $q=1.04$ and $\beq/\aeq=0.0$ for $N=1$, $5$, $10$, $15$, and $20$.
Figure~\ref{fig:StTwo:rescaled} shows the Tsallis entropies divided by $N$, $\STtwo/N$,
as functions of $t^{(2)}$ at $q=1.04$ and $\beq/\aeq=0.0$ for $N=1$, $5$, $10$, $15$, and $20$.
The Tsallis entropy per oscillator decreases with $N$ at high scaled temperature,
while the Tsallis entropy per oscillator increases with $N$ at low scaled temperature.
A fixed point does not exist in Fig.~\ref{fig:StTwo:rescaled}, even though it seems that the lines intersect at the same point.
Figure~\ref{fig:StTwo:qdep} shows the Tsallis entropies $\STtwo$ as functions of $t^{(2)}$
at $N=20$ and $\beq/\aeq=0.0$ for $q=1.01$, $1.02$, $1.03$, and $1.04$. 
The Tsallis entropy decreases with $q$ at high scaled temperature,
while the Tsallis entropy increases with $q$ at low scaled temperature.
A fixed point does not also exist in Fig.~\ref{fig:StTwo:qdep}.  
Figure~\ref{fig:StTwo:bdep} shows the Tsallis entropies $\STtwo$
as functions of $t^{(2)}$ at $q=1.04$ and $N=20$ for $\beq/\aeq = 0.0$, $0.5$, $1.0$, and $1.5$. 
The Tsallis entropies approach non-zero values as $t^{(2)}$ goes to zero, except when $\beq/\aeq =0.0$.
This behavior can be explained from Eq.~\eqref{eqn:Tsallis2:gammatwo}.
The value of $\gammatwo$ approaches a positive value for $\beq/\aeq > 0$,
while the value of $\gammatwo$ approaches zero for $\beq/\aeq=0$, as $\scTtwo$ goes to zero. 
Therefore, the Tsallis entropy $\STtwo$ approaches a positive value for $\beq/\aeq>0$ as $\scTtwo$ goes to zero.


\begin{figure}
  \centering
  \subfigure[The heat capacities] 
            {
              \includegraphics[width=0.45\textwidth]{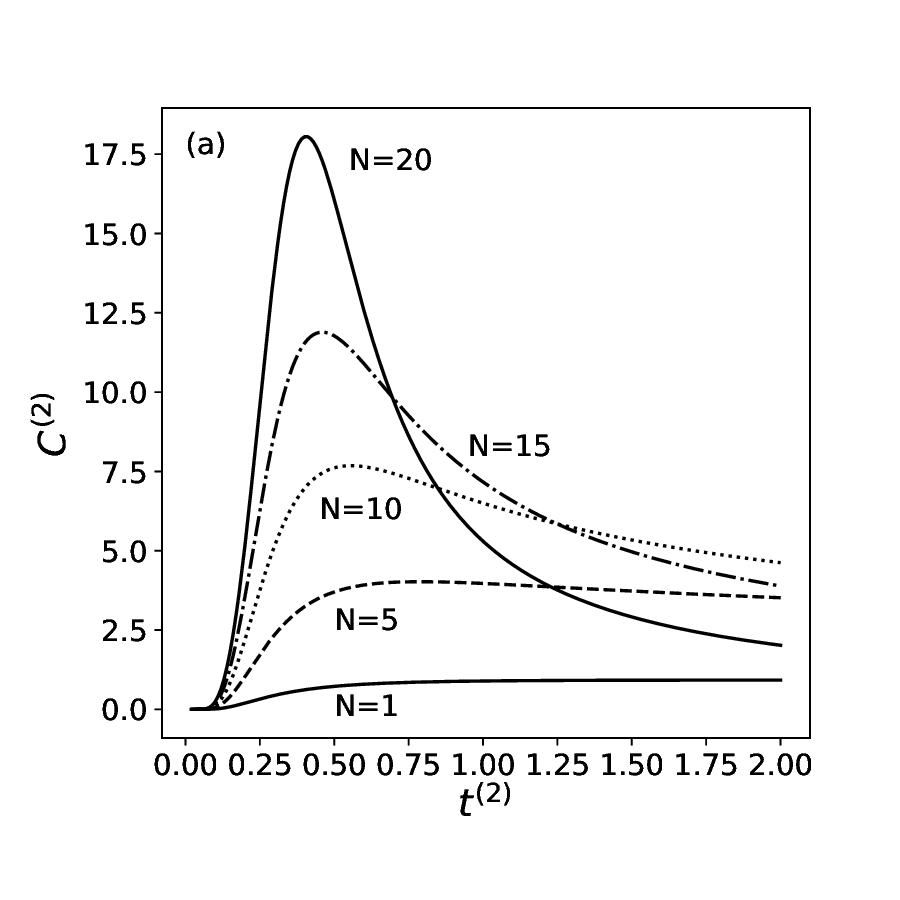}
              \label{fig:2:Cv:Ndep}
            }
            \hfill
            \subfigure[The heat capacities divided by $N$]
                      {
            \includegraphics[width=0.45\textwidth]{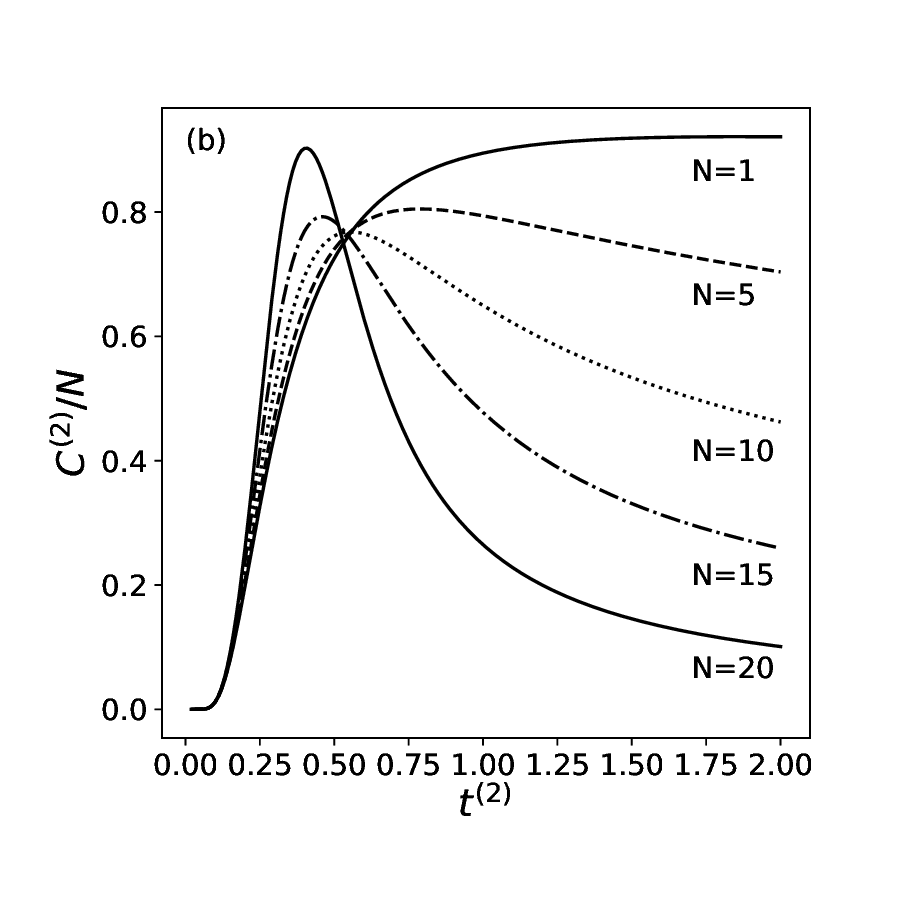}
            \label{fig:2:Cv:Ndep:b}
            }
              \caption{
                  The heat capacities $\CvCEtwo$ and the heat capacities divided by $N$, $\CvCEtwo/N$,
                  as functions of the scaled temperature $t^{(2)}$ at $q=1.04$ and $\beq/\aeq=0.0$ for $N=1$, $5$, $10$, $15$, and $20$.  }
\end{figure}

\begin{figure}[tbp]
  \begin{minipage}[b]{0.48\columnwidth}
    \centering
    \includegraphics[width=0.94\textwidth]{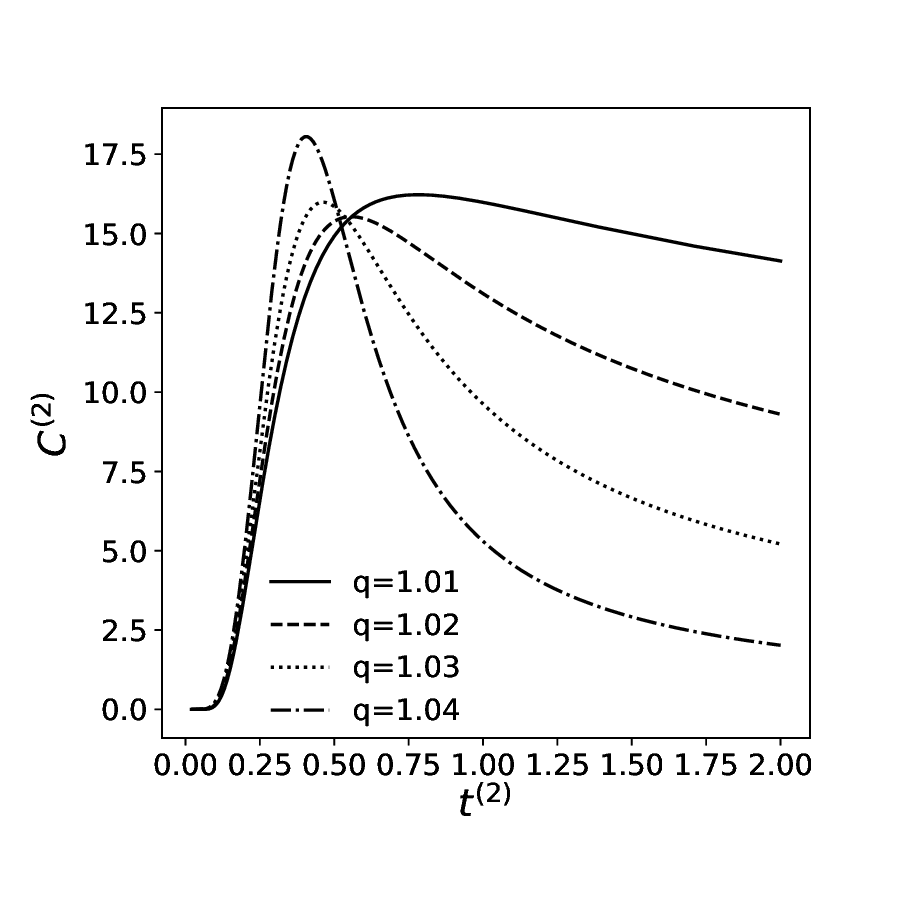}
    \caption{The heat capacities $\CvCEtwo$ as functions of the scaled temperature $t^{(2)}$
      at $N=20$ and $\beq/\aeq=0.0$ for $q=1.01$, $1.02$, $1.03$ and $1.04$.}
  \label{fig:2:Cv:qdep}
  \end{minipage}
  \hspace{0.04\columnwidth} 
  \begin{minipage}[b]{0.48\columnwidth}
  \centering
  \includegraphics[width=0.94\textwidth]{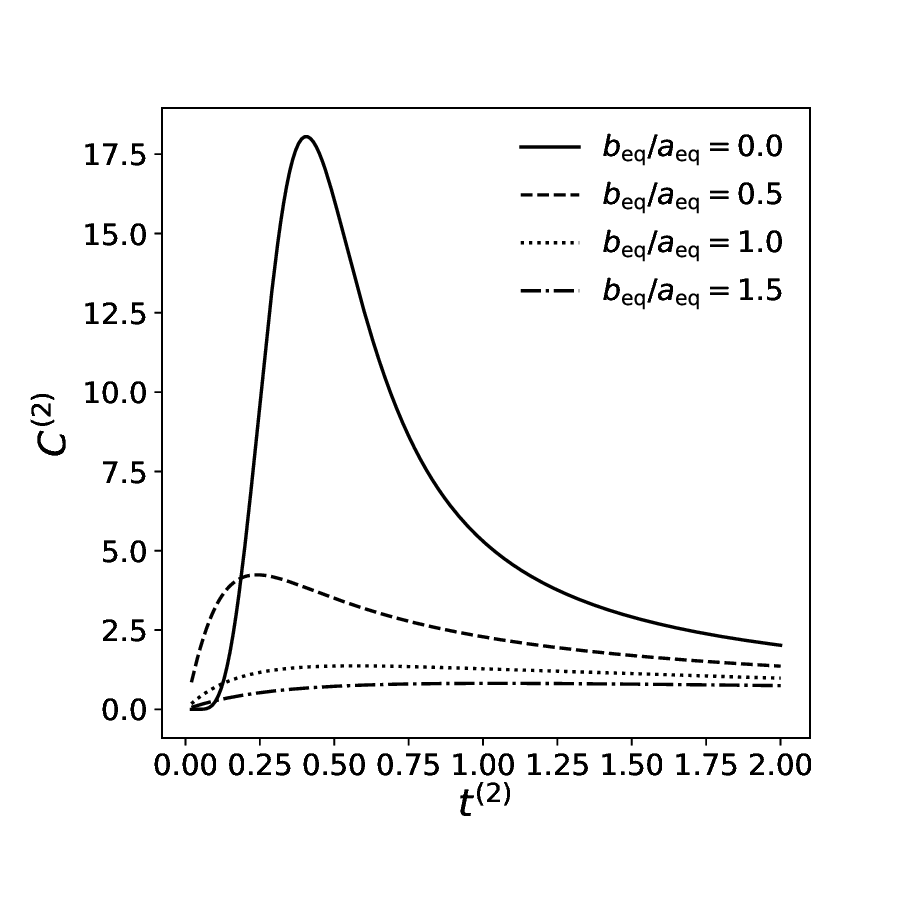}
  \caption{The heat capacities $\CvCEtwo$ as functions of the scaled temperature $t^{(2)}$
    at $q=1.04$ and $N=20$ for $\beq/\aeq = 0.0$, $0.5$, $1.0$ and $1.5$.
  }
  \label{fig:2:Cv:bdep}
  \end{minipage}
\end{figure}

Finally, we calculate the heat capacity numerically. 
Figure~\ref{fig:2:Cv:Ndep} shows the heat capacities $\CvCEtwo$
as functions of $t^{(2)}$ at $q=1.04$ and $\beq/\aeq = 0.0$ for $N=1$, $5$, $10$, $15$, and $20$.
The behavior of the heat capacity in the Tsallis-2 statistics is quite different from that in the Boltzmann-Gibbs statistics.
As shown in this figure, the heat capacity increases with the scaled temperature, and reaches the maximum, and decreases after that.
This behavior reflects the scaled temperature dependence of the energy. 
Figure~\ref{fig:2:Cv:Ndep:b} shows the heat capacities divided by $N$, $\CvCEtwo/N$, 
as functions of $t^{(2)}$ at $q=1.04$ and $\beq/\aeq = 0.0$ for $N=1$, $5$, $10$, $15$, and $20$.
The heat capacity per oscillator, $\CvCEtwo/N$, decreases with $N$ at high scaled temperature,
while the heat capacity per oscillator increases with $N$ at low scaled temperature.
A fixed point does not exist in Fig.~\ref{fig:2:Cv:Ndep:b}, even though it seems that the lines intersect at the same point.
Figure~\ref{fig:2:Cv:qdep} shows the heat capacities $\CvCEtwo$
as functions of $t^{(2)}$ at $N=20$ and $\beq/\aeq = 0.0$ for $q=1.01$, $1.02$, $1.03$ and $1.04$. 
The heat capacity decreases with $q$ at high scaled temperature,
while the heat capacity increases with $q$ at low scaled temperature.
A fixed point does not also exist in Fig.~\ref{fig:2:Cv:qdep}.
Figure~\ref{fig:2:Cv:bdep} shows the heat capacities $\CvCEtwo$
as functions of $t^{(2)}$ at $q=1.04$ and $N=20$ for $\beq/\aeq = 0.0$, $0.5$, $1.0$, and $1.5$. 
The heat capacity depends on $\beq$, because the energy depends on $\beq$.
The variation of the heat capacity is weak for large $\beq/\aeq$.

\section{Multiple quantum harmonic oscillators in the Tsallis statistics with the normalized $q$-expectation value }
\label{Sec:Tsallis3}

In this section,
we deal with the energy, the Tsallis entropy, the average level of the oscillators, and the heat capacity 
for multiple quantum harmonic oscillators in the Tsallis-3 statistics:
we employ the Tsallis entropy and the normalized $q$-expectation value.
The probability in the Tsallis-3 statistics is invariant to energy shift. 
The normalized $q$-expectation value is often called the escort average which satisfies the property $\Eq{\hat{1}}=1$.
We calculate the energy, the Tsallis entropy, and the heat capacity numerically.

\subsection{Physical quantities for multiple quantum harmonic oscillators in the Tsallis statistics with the normalized $q$-expectation value }


We attempt to obtain the probability $p^{(3)}$, the energy $U^{(3)}$, and the Tsallis entropy $\STthree$
by using Eqs.~\eqref{Tsallis-3:density_op} and \eqref{Tsallis-3:expectation} for the energy, Eq.~\eqref{eqn:En}.
The energy $U^{(3)}$ is calculated by
\begin{align}
  U^{(3)} = \frac{\displaystyle\sum_{\{n\}} \big( p^{(3)}(\{n\}) \big)^q E(\{n\})}{\displaystyle\sum_{\{n\}} \big( p^{(3)}(\{n\}) \big)^q}
  = \frac{\displaystyle\sum_{n_1=0,\cdots,n_N=0}^{\infty} \big( p^{(3)}(\{n\}) \big)^q
    \Bigg(\displaystyle\sum_{j=1}^N (a_j n_j + b_j)\Bigg)}{\displaystyle\sum_{n_1=0,\cdots,n_N=0}^{\infty}\big( p^{(3)}(\{n\}) \big)^q} .
\end{align}
Here we define $\ER$ and $\URthree$ by
\begin{align}
  &\ER = E - \sum_{j=1}^{N} b_j ,\\
  &\URthree =\Uthree - \sum_{j=1}^{N} b_j .
\end{align}
Evidently, $\URthree$ is represented by
\begin{align}
\URthree = \frac{\displaystyle\sum_{\{n\}} (p^{(3)})^q \ER}{\displaystyle\sum_{\{n\}}(p^{(3)})^q} . 
\end{align}

We obtain the probability $p^{(3)}$, the energy $\URthree$, and the Tsallis entropy $\STthree$: 
\begin{subequations}
\begin{align}
  & p^{(3)} = \frac{\Bigg[  \alphathree + \displaystyle\sum_{j=1}^N a_j n_j \Bigg]^{1/(1-q)}}{\Bzeta{1/(q-1)}{\alphathree}{\vec{a}_N}} ,\\
  &\URthree = \frac{\Bzeta{1/(q-1)}{\alphathree}{\vec{a}_N}}{\Bzeta{q/(q-1)}{\alphathree}{\vec{a}_N}} - \alphathree , \label{eqn:UR}\\
  &\STthree = (q-1)^{-1} \left\{ 1 - \frac{\Bzeta{q/(q-1)}{\alphathree}{\vec{a}_N}}{\left[\Bzeta{1/(q-1)}{\alphathree}{\vec{a}_N}\right]^q} \right\} ,
\end{align}
\end{subequations}
where $\alphathree$ is given by
\begin{align}
\alphathree = \frac{1+(1-q)\betaPh \URthree}{(q-1)\betaPh}.
\label{def:alphathree}
\end{align}
We have the limitation $1 < q < (N+1)/N$ in the Tsallis-3 statistics, as in the Tsallis-2 statistics.

In the Tsallis-3 statistics, there is the relation $c_q = (Z^{(3)})^{1-q}$,
where the partition function $Z^{(3)}$ in the present case is given by
\begin{align}
Z^{(3)} = [(q-1) \betaPh]^{1/(1-q)} \Bzeta{1/(q-1)}{\alphathree}{\vec{a}_N} . 
\end{align}
This relation  $c_q = (Z^{(3)})^{1-q}$ gives the self-consistent equation:
\begin{align}
  \Bzeta{q/(q-1)}{\alphathree}{\vec{a}_N} = (q-1) \betaPh \Bzeta{1/(q-1)}{\alphathree}{\vec{a}_N} .
\label{eqn:self-consistent}
\end{align}
Equation~\eqref{eqn:self-consistent} is also derived from Eqs.~\eqref{eqn:UR} and \eqref{def:alphathree}.
The self-consistent equation, Eq.~\eqref{eqn:self-consistent}, is consistent with the expression of the energy.

We now focus on the case where $a_1$, $a_2$, $\cdots$, $a_N$ are all equal to $\aeq$.
It is worth to mention that no condition for $b_j$ is imposed. 
We introduce the following quantity $\tPh$ and $\gammathree$ by 
\begin{subequations}
\begin{align}
&\tPh = \frac{1}{\aeq \betaPh} , \\
&\gammathree=\frac{\alphathree}{\aeq} . 
\end{align}
\end{subequations}
The average level $\numberthree$ is defined by
\begin{align}
  \numberthree
  = \frac{1}{N}
  \frac{\displaystyle\sum_{n_1=0, \cdots, n_N=0}^{\infty} \left( \sum_{j=1}^N n_j \right) \left( p^{(3)} \right)^q}
       {\displaystyle\sum_{n_1=0, \cdots, n_N=0}^{\infty} \left( p^{(3)} \right)^q}.
\end{align}
The heat capacity $\CvCEthreePh$ is defined by 
\begin{align}
  \CvCEthreePh  = \frac{\partial \Uthree}{\partial \TPh},
\end{align}
where $\TPh$ is the equilibrium temperature (the physical temperature): $\TPh$ is given by $1/\betaPh$.

The probability $p^{(3)}$, the scaled energy $\URthree/\aeq$, the Tsallis entropy $\STthree$,
the average level $\numberthree$, and the heat capacity $\CvCEthreePh$  are represented by
\begin{subequations}
  \begin{align}
    &p^{(3)} = \frac{ \Bigg[ \gammathree + \displaystyle\sum_{j=1}^N n_j \Bigg]^{1/(1-q)}}{\Bzeta{1/(q-1)}{\gammathree}{\vec{1}_N}} , \\
    &\frac{\URthree}{\aeq} = \frac{\Bzeta{1/(q-1)}{\gammathree}{\vec{1}_N}}{\Bzeta{q/(q-1)}{\gammathree}{\vec{1}_N}} - \gammathree , \label{Tsallis3:u}\\
    &\STthree = \frac{1}{(q-1)}
    \left\{ 1 - \frac{\Bzeta{q/(q-1)}{\gammathree}{\vec{1}_N}}{\Bigg[\Bzeta{1/(q-1)}{\gammathree}{\vec{1}_N}\Bigg]^{q}}\right\} , \\
    &\numberthree = \frac{1}{N} \left\{ \frac{\Bzeta{1/(q-1)}{\gammathree}{\vec{1}_N}}{\Bzeta{q/(q-1)}{\gammathree}{\vec{1}_N}} - \gammathree \right\} ,  \label{Tsallis3:n}\\
    &\CvCEthreePh
  = \frac{\partial \URthree}{\partial \TPh} = \frac{\partial (\URthree/\aeq)}{\partial \tPh} 
  = \frac{1}{(q-1)}
  \left(
  \frac{\frac{\partial}{\partial \gammathree} \left(\URthree/\aeq \right)}
       {1 + \frac{\partial}{\partial \gammathree} \left(\URthree/\aeq \right)}
  \right) \nonumber \\
  & \quad\quad = \frac{1}{(q-1)}
  + \frac{\left[\Bzeta{q/(q-1)}{\gammathree}{\vec{1}_N}\right]^2}
  {\left[\Bzeta{q/(q-1)}{\gammathree}{\vec{1}_N}\right]^2 - q \Bzeta{1/(q-1)}{\gammathree}{\vec{1}_N} \Bzeta{(2q-1)/(q-1)}{\gammathree}{\vec{1}_N} } .
  \end{align}
\end{subequations}
The self-consistent equation, Eq.~\eqref{eqn:self-consistent}, is represented as follows:
\begin{align}
\tPh \Bzeta{q/(q-1)}{\gammathree}{\vec{1}_N} = (q-1) \Bzeta{1/(q-1)}{\gammathree}{\vec{1}_N}  .
\end{align}

We find the relation $\URthree = N \aeq \numberthree$ from Eqs.~\eqref{Tsallis3:u} and \eqref{Tsallis3:n}.
That is 
\begin{align}
\Uthree = N \aeq \numberthree + \sum_{j=1}^N b_j .
\label{eqn:Uthree:numberthree}
\end{align}
The last term is the zero-point energy of the oscillators.
Equation~\eqref{eqn:Uthree:numberthree} is directly derived from the definitions of $\Uthree$ and $\numberthree$:
\begin{align}
  \Uthree
  &= \frac{\displaystyle \sum_{\{n\}} \Big( \sum_{j=1}^N (\aeq n_j + b_j) \Big) \big(p(\{n\})\big)^q}{\displaystyle \sum_{\{n\}}  \big(p(\{n\})\big)^q}
  = \aeq \frac{\displaystyle \sum_{\{n\}} \Big( \sum_{j=1}^N n_j \Big) \big(p(\{n\})\big)^q}{\displaystyle \sum_{\{n\}} \big(p(\{n\})\big)^q}    
  + \sum_{j=1}^N b_j
  = N \aeq \numberthree + \sum_{j=1}^N b_j.
\end{align}
The simple relation between $\Uthree$ and $\numberthree$ is obtained in the case that $b_1$, $\cdots$, $b_N$ are all equal to $\beq$:
\begin{align}
  \Uthree  = N (\aeq \numberthree + \beq). 
\end{align}
The natural relation between the energy and the average level holds in the Tsallis-3 statistics.

\subsection{Numerical results in the Tsallis statistics with the normalized $q$-expectation value}
In the Tsallis statistics with the normalized $q$-expectation value,
we treat the case where $a_1$, $a_2$, $\cdots$, $a_N$ are all equal to $\aeq$.
We calculate the scaled energy $\URthree/\aeq$, the Tsallis entropy $\STthree$, and the heat capacity $\CvCEthreePh$
as functions of the scaled equilibrium temperature (the scaled physical temperature) $\tPh$ numerically. 
The value of $q$ is chosen to satisfy the inequality $1 < q < (N+1)/N$, as chosen in the previous section. 

\begin{figure}
  \centering
  \subfigure[The scaled energies]
            {\includegraphics[width=0.4\textwidth]{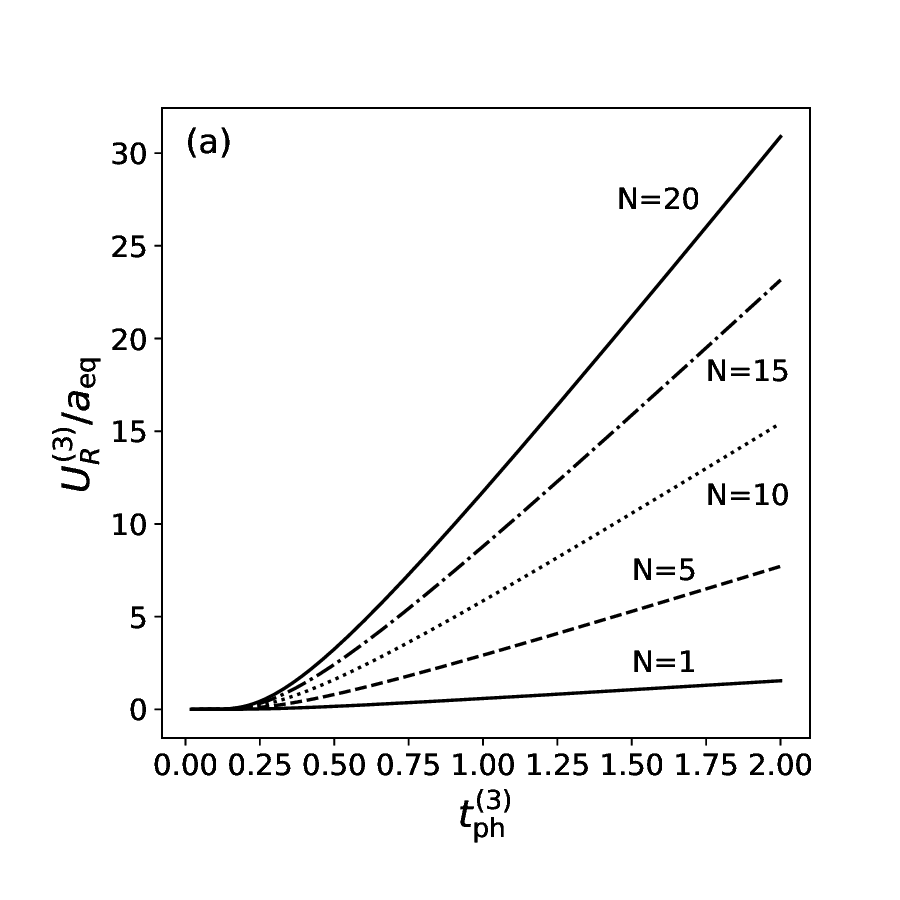}
            \label{fig:3:UR:a}
            }
            \hfill
  \subfigure[The scaled energies divided by $N$]
            {\includegraphics[width=0.4\textwidth]{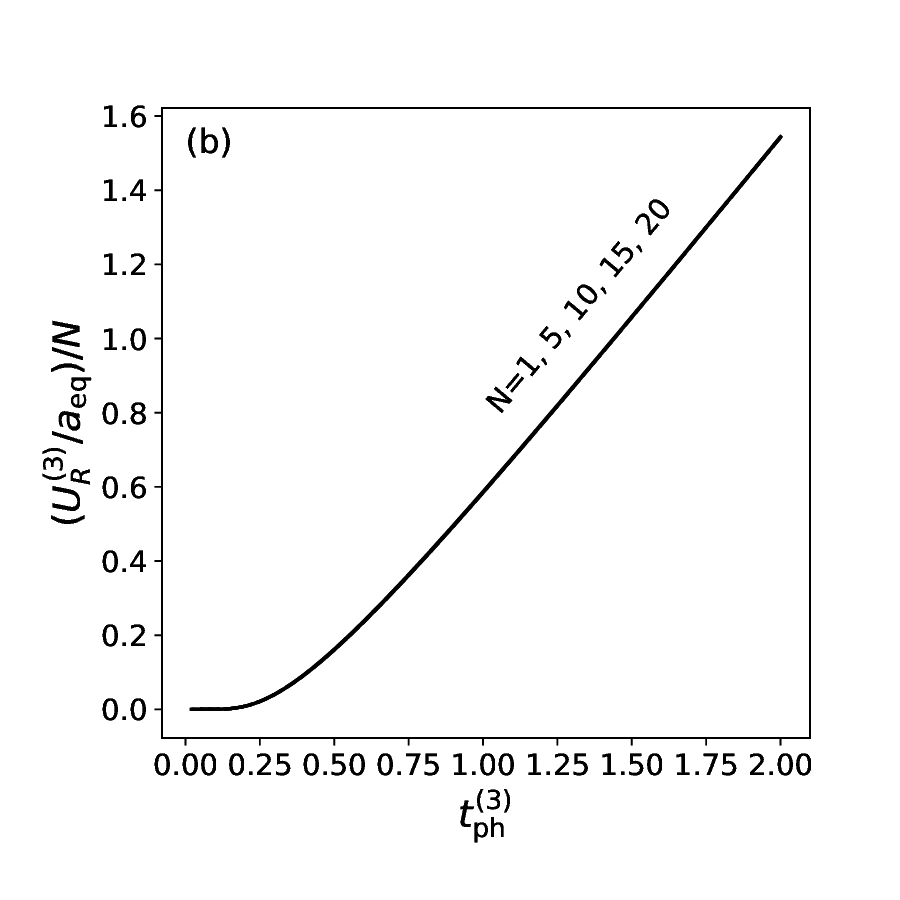}
            \label{fig:3:UR:b}
            }
  \subfigure[The scaled energies divided by $N$ in the narrow range]
            {\includegraphics[width=0.4\textwidth]{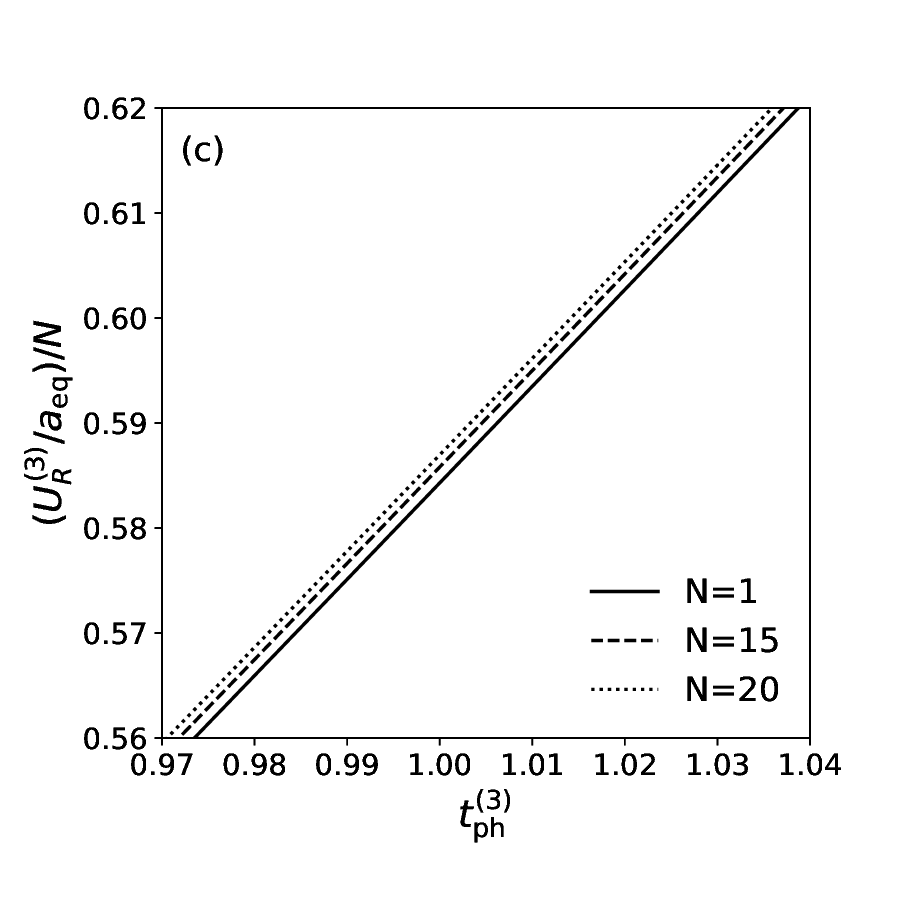}
            \label{fig:3:UR:c}
            }
            \caption{
              The scaled energies $\URthree/\aeq$, the scaled energies divided by $N$, $(\URthree/\aeq)/N$,
              and the scaled energies divided by $N$ in the narrow range,  
              as functions of the scaled equilibrium temperature (the scaled physical temperature) $\tPh$ at $q=1.03$.
              The values of $N$ are $1, 5, 10, 15$, and $20$ in Figs.~\ref{fig:3:UR:a} and \ref{fig:3:UR:b},
              while the values of $N$ are $1, 15$, and $20$ in Fig.~\ref{fig:3:UR:c}.
            }
  \label{fig:3:UR}
\end{figure}

\begin{figure}
  \centering
  \subfigure[The scaled energies]
            {\includegraphics[width=0.4\textwidth]{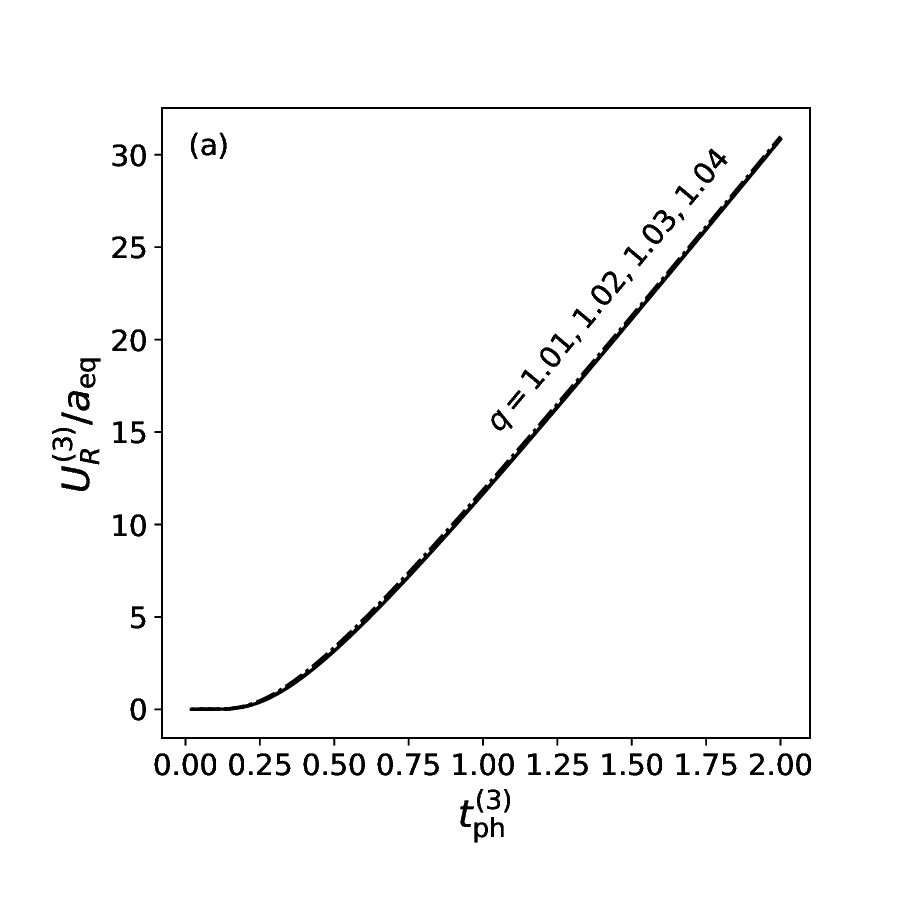}
              \label{fig:3:UR:qdep}
            }
            \hfill
  \subfigure[The log-log plot of the scaled energies]
            {\includegraphics[width=0.4\textwidth]{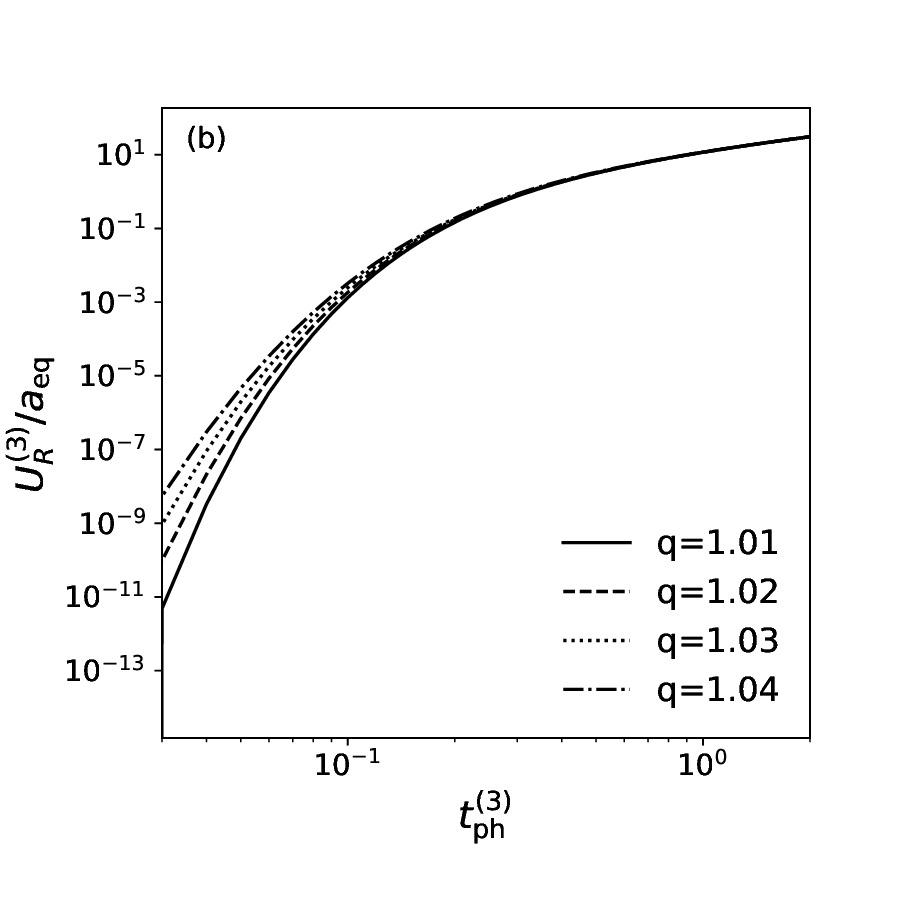}
              \label{fig:3:UR:qdep:log-log}
            }
  \caption{The scaled energies $\URthree/\aeq$ as functions of the scaled equilibrium temperature
    (the scaled physical temperature) $\tPh$ at $N=20$ for $q=1.01, 1.02, 1.03$, and $1.04$.}
  \label{fig:3:UR:qdep:all}
\end{figure}

First, we calculate the scaled energy numerically. 
Figure~\ref{fig:3:UR:a} shows the scaled energies $\URthree/\aeq$ as functions of $\tPh$ at $q=1.03$
for $N=1$, $5$, $10$, $15$, and $20$.
Figure.~\ref{fig:3:UR:b} shows the scaled energies divided by $N$, $(\URthree/\aeq)/N$, as functions of $\tPh$ at $q=1.03$
for $N=1$, $5$, $10$, $15$, and $20$.
Figure.~\ref{fig:3:UR:c} shows the scaled energies divided by $N$ as functions of $\tPh$ at $q=1.03$
for $N=1$, $15$, and $20$ in the narrow range.
The scaled energy $\URthree/\aeq$ increases with $\tPh$,
and the $N$ dependence of the scaled energy per oscillator, $(\URthree/\aeq)/N$, is exceedingly weak.   
The $N$ dependence is seen in Fig.~\ref{fig:3:UR:c}.
The energy for $N$ oscillators is approximately $N$ times the energy for a single oscillator.
Figure~\ref{fig:3:UR:qdep} and Fig.~\ref{fig:3:UR:qdep:log-log}
show the scaled energies $\URthree/\aeq$ as functions of $\tPh$ at $N=20$ for $q=1.01, 1.02, 1.03$, and $1.04$. 
Figure~\ref{fig:3:UR:qdep:log-log} is the log-log plot of the scaled energies.
The $q$ dependence of $\URthree/\aeq$ is quite weak.
The dependence is shown in Fig.~\ref{fig:3:UR:qdep:log-log} explicitly.

\begin{figure}
  \centering
  \subfigure[The Tsallis entropies]
            {\includegraphics[width=0.4\textwidth]{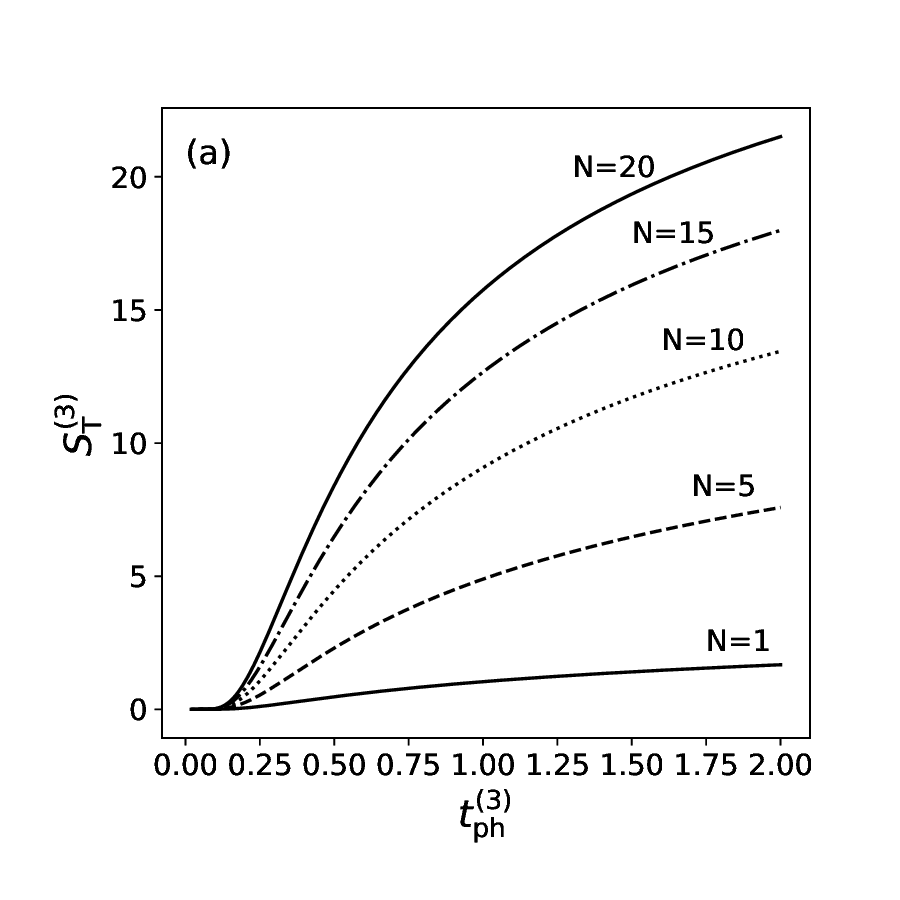}
            \label{fig:3:S:a}
            }
            \hfill
            \subfigure[The Tsallis entropies divided by $N$]
            {\includegraphics[width=0.4\textwidth]{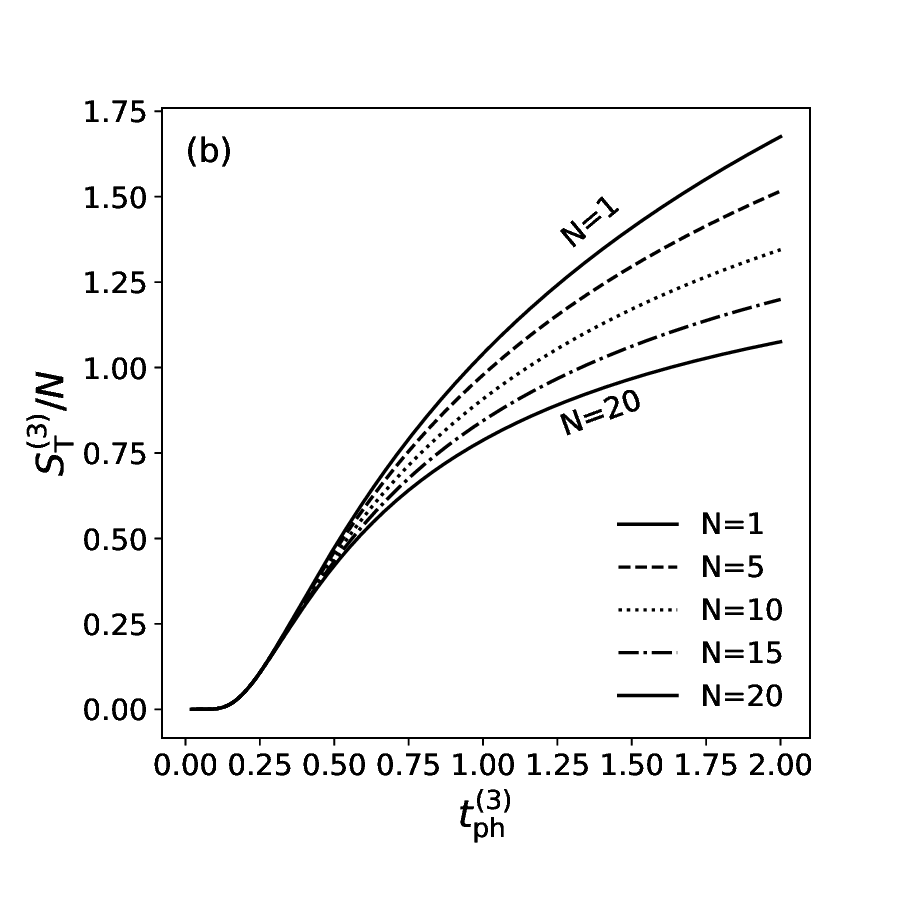}
            \label{fig:3:S:b}
            }
            \caption{The Tsallis entropies $\STthree$ as functions of the scaled equilibrium temperature (the scaled physical temperature) $\tPh$
              at $q=1.03$ for $N=1, 5, 10, 15$, and $20$.}
  \label{fig:3:S}
\end{figure}

\begin{figure}
  \centering
  \includegraphics[width=0.4\textwidth]{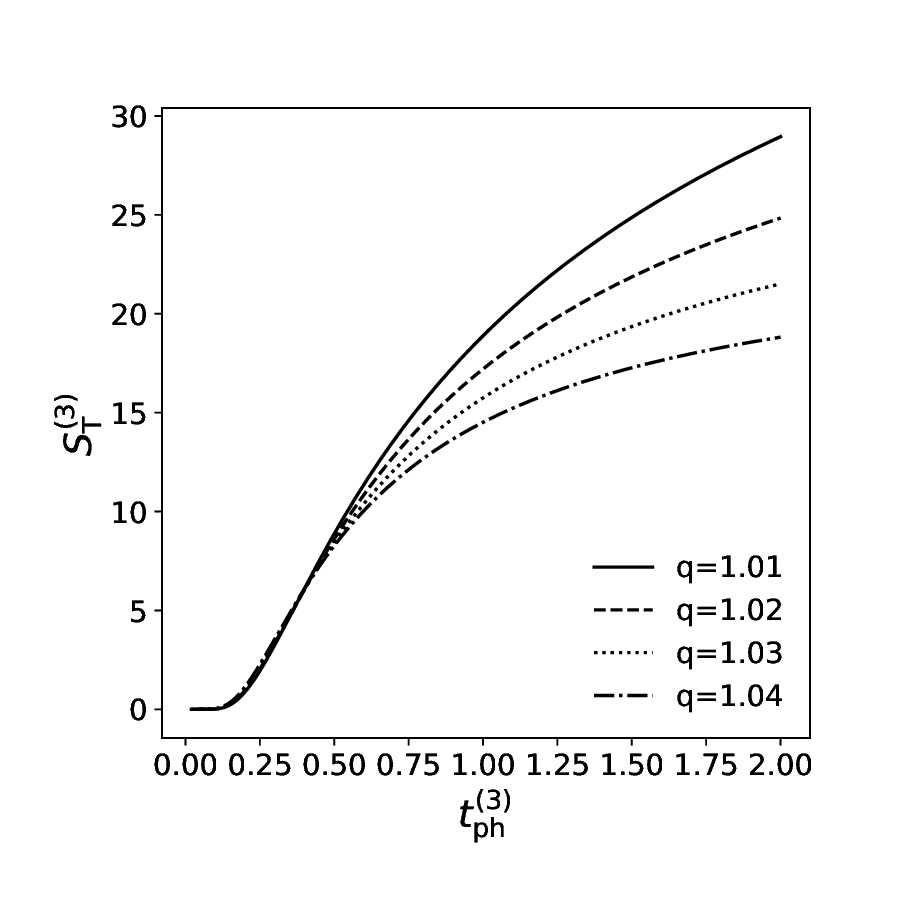}
  \caption{The Tsallis entropies $\STthree$ as functions of the scaled equilibrium temperature (the scaled physical temperature) $\tPh$
    at $N=20$ for $q=1.01$, $1.02$, $1.03$, and $1.04$.}
  \label{fig:3:S:qdep}
\end{figure}

Next, we calculate the Tsallis entropy numerically. 
Figure~\ref{fig:3:S:a} shows the Tsallis entropies $\STthree$ as functions of $\tPh$ at $q=1.03$ for $N=1, 5, 10, 15$ and $20$.
Figure~\ref{fig:3:S:b} shows the Tsallis entropies divided by $N$, $\STthree/N$, as functions of $\tPh$ at $q=1.03$ for $N=1, 5, 10, 15$ and $20$.
The Tsallis entropy $\STthree$ increases with $\tPh$.
The Tsallis entropy per single oscillator decreases with $N$. 
Figure~\ref{fig:3:S:qdep} shows the Tsallis entropies $\STthree$ as functions of $\tPh$ at $N=20$
for $q=1.01$, $1.02$, $1.03$, and $1.04$.
The Tsallis entropy decreases with $q$.

\begin{figure}
  \centering
  \subfigure[The heat capacities] 
            {\includegraphics[width=0.4\textwidth]{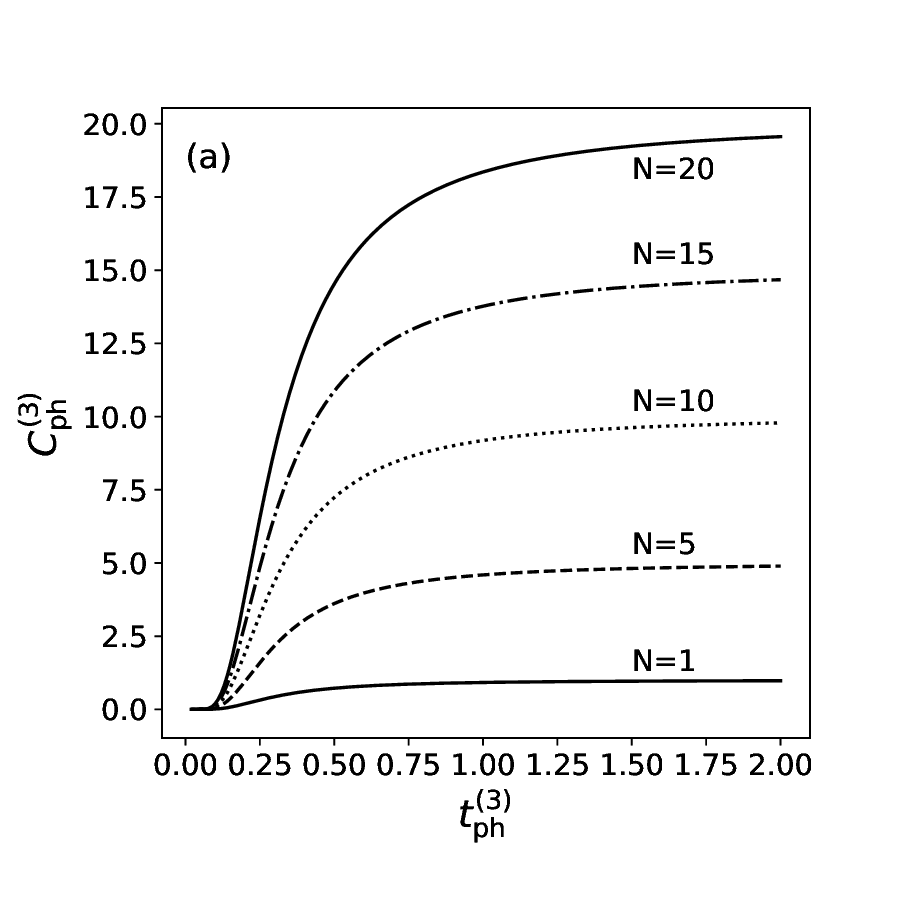}
              \label{fig:3:C:a}
            }
            \hfill
            \subfigure[The heat capacities divided by $N$]
            {
              \includegraphics[width=0.4\textwidth]{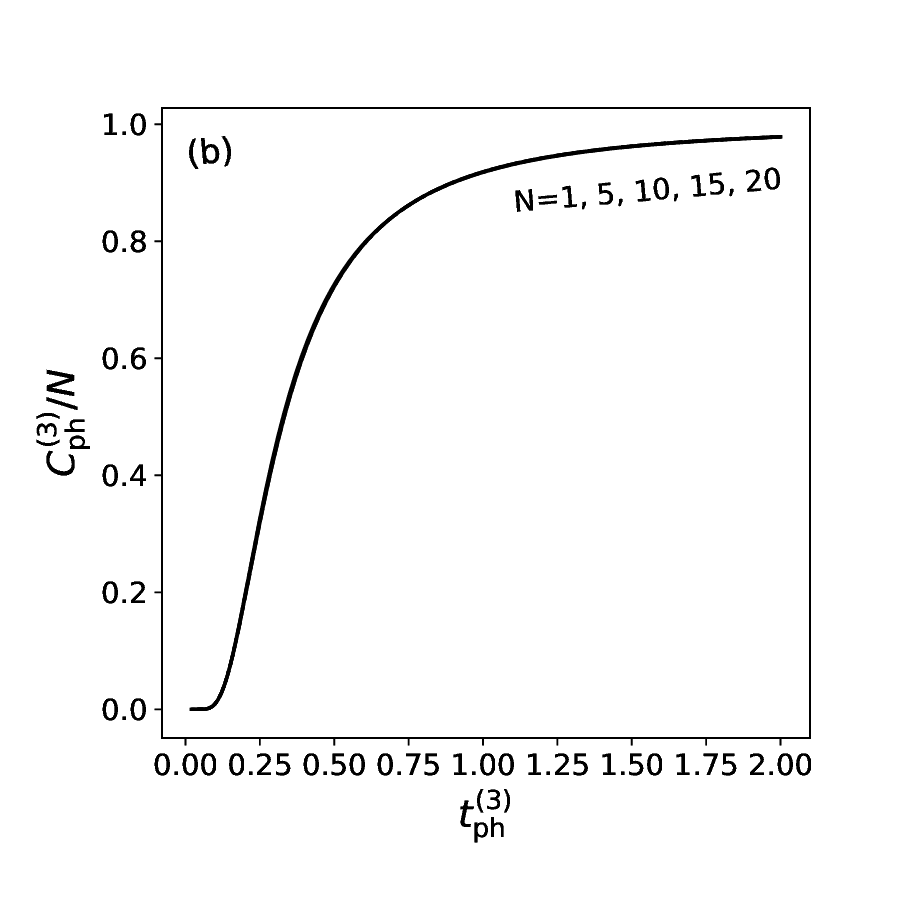}
              \label{fig:3:C:b}
            }
            \caption{The heat capacities $\CvCEthreePh$ and the heat capacities divided by $N$, $\CvCEthreePh/N$,
              as functions of the scaled equilibrium temperature (the scaled physical temperature) $\tPh$ at $q=1.03$ for $N=1, 5, 10, 15$, and $20$.}
  \label{fig:3:C}
\end{figure}

\begin{figure}
  \centering
  \includegraphics[width=0.4\textwidth]{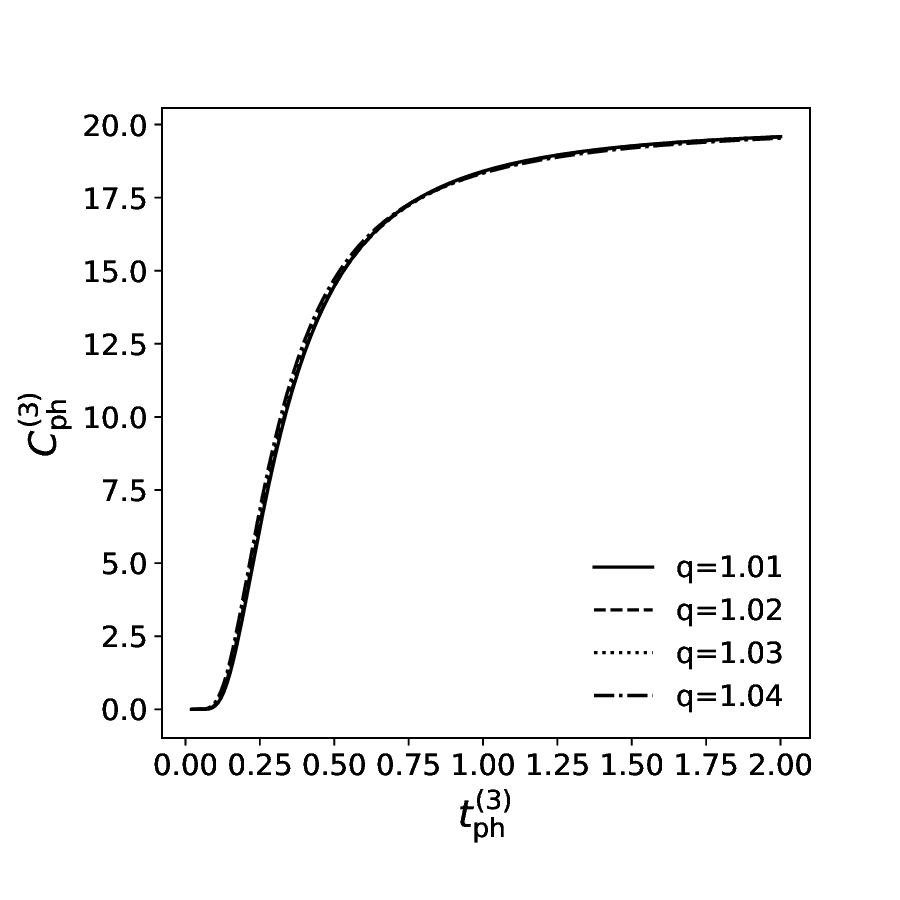}
  \caption{The heat capacities $\CvCEthreePh$ as functions of the scaled equilibrium temperature
    (the scaled physical temperature) $\tPh$ at $N=20$ for $q=1.01$, $1.02$, $1.03$, and $1.04$.}
  \label{fig:3:C:qdep}
\end{figure}

Finally, we calculate the heat capacity numerically. 
Figure~\ref{fig:3:C:a} shows the heat capacities $\CvCEthreePh$ as functions of $\tPh$ at $q=1.03$
for $N=1$, $5$, $10$, $15$, and $20$.
Figure~\ref{fig:3:C:b} shows the heat capacities divided by $N$, $\CvCEthreePh/N$, as functions of $\tPh$ at $q=1.03$
for $N=1$, $5$, $10$, $15$, and $20$.
The difference in the heat capacity divided by $N$ cannot be seen explicitly in Fig.~\ref{fig:3:C:b}.
The $N$ dependence of the heat capacity is exceedingly weak.
This is a direct result of the $\tPh$ dependence of the energy divided by $N$ as shown in Fig.~\ref{fig:3:UR:b}.
Therefore, the heat capacity $\CvCEthreePh$ for $N$ oscillators is approximately   
$N$ times the heat capacity $\CvCEthreePh$ for a single oscillator.
Figure~\ref{fig:3:C:qdep} shows the heat capacities $\CvCEthreePh$
as functions of $\tPh$ at $N=20$ for $q = 1.01, 1.02, 1.03$, and $1.04$.
The behavior of the heat capacity reflects the behavior of the energy.
Therefore, the $q$ dependence of the heat capacity is quite weak. 

\section{Discussions and conclusions}
\label{Sec:Conclusions}
We studied the thermodynamic quantities of the system whose energy is represented as $\displaystyle\sum_{j=1}^N (a_j n_j + b_j)$
in the Tsallis statistics of entropic parameter $q$, where $N$ is the number of oscillators, in the case that the distribution is power-like.
We employed 
the Tsallis-1 statistics in which the expectation value is the conventional expectation value, 
the Tsallis-2 statistics in which the expectation value is the unnormalized $q$-expectation value,
and the Tsallis-3 statistics in which the expectation value is the normalized $q$-expectation value  (the escort average).
We obtained the expressions of the energy and the Tsallis entropy.
We also obtain the expressions of the energy, the Tsallis entropy,
the average level of the oscillators, and the heat capacity
for $a_1=\cdots=a_N$ in the Tsallis-1 statistics and the Tsallis-3 statistics. 
These expressions are also obtained for $a_1=\cdots=a_N$ and $b_1=\cdots=b_N$ in the Tsallis-2 statistics.
Numerically, we calculated the energy, the Tsallis entropy, and the heat capacity  
for $a_1=\cdots=a_N$  in the Tsallis-1 statistics and the Tsallis-3 statistics. 
We also calculated these quantities for $a_1=\cdots=a_N$ and $b_1=\cdots=b_N$ in the Tsallis-2 statistics.
These quantities were calculated using the expansion of the Barnes zeta function with the Hurwitz zeta function. 

The parameter $q$ is less than one for the power-like distributions in the Tsallis-1 statistics. 
This parameter $q$ is greater than one for the power-like distributions in the Tsallis-2 statistics and the Tsallis-3 statistics.
The physical quantities calculated in this paper are represented with the Barnes zeta function in these statistics.
The Barnes zeta function requires the condition $q/(1-q) > N$ for the Tsallis-1 statistics.
The Barnes zeta function also requires the condition $1/(q-1) > N$ for the Tsallis-2 statistics and the Tsallis-3 statistics.
The quantity $|q-1|$ is the measure of the deviation from the Boltzmann-Gibbs statistics.
In the Tsallis-1 statistics, there is the simple relation between the energy and the average level in the present system.
The probability is represented with the energy and the Tsallis entropy, and does not contain the zero-point energy.
In the Tsallis-2 statistics, there is the relation among the energy, the average level, and the Tsallis entropy in the present system. 
The zero-point energy might play significant roles in the Tsallis-2 statistics, because the zero-point energy appears explicitly.
In the Tsallis-3 statistics, there is the simple relation between the energy and the average level in the present system.
The physical quantities except for the energy are not affected by the zero-point energy, 
when the equilibrium temperature (the physical temperature) is adopted.
The self-consistent equation coming from the energy constraint is the same self-consistent equation coming from the normalization condition.

In the Tsallis-1 statistics, the behaviors were clarified from the numerical calculations
in the system of oscillators with identical energy difference: $a_1 = a_2 = \cdots = a_N$.
The physical quantities as functions of the scaled temperature $t^{(1)}$,
which is the temperature divided by the difference between adjacent energy levels, depend on $N$ and $q$.
The energy, the Tsallis entropy, and the heat capacity were studied.
These quantities divided by $N$ increase with $t^{(1)}$, and increase with $N$.
The energy, the Tsallis entropy, and the heat capacity decrease with $q$. 
These results indicate that the quantity for $N$ independent oscillators is not $N$ times the quantity for a single oscillator.

In the Tsallis-2 statistics, the behaviors were clarified from the numerical calculations
in the system of oscillators with identical energy levels: $a_1 = a_2 = \cdots = a_N$ and $b_1 = b_2 = \cdots = b_N$.
The physical quantities as functions of the scaled temperature $t^{(2)}$ depend on $N$ and $q$.
The energy, the Tsallis entropy, and the heat capacity were also studied.
These quantities divided by $N$ increase with $N$ at low scaled temperature, 
while decrease with $N$ at high scaled temperature.
The energy, the Tsallis entropy, and the heat capacity increase with $q$ at low scaled temperature, while decrease with $q$ at high scaled temperature.
These results indicate that the quantity for $N$ independent oscillators is not $N$ times the quantity for a single oscillator.
In the Tsallis-2 statistics,
the heat capacity has a peak as a function of $\scTtwo$. 
The heat capacity resembles the Schottky-type heat capacity.

In the Tsallis-3 statistics, the behaviors were clarified from the numerical calculations
in the system of oscillators with identical energy difference: $a_1 = a_2 = \cdots = a_N$.
We employed the equilibrium temperature (the physical temperature) to describe the quantities.
The $N$ dependence of the energy divided by $N$ is exceedingly weak. 
The $N$ dependence of the heat capacity divided by $N$ is also exceedingly weak,
where the heat capacity is defined by the derivative of the energy with respect to the equilibrium temperature. 
The energy and the heat capacity for $N$-independent oscillators are approximately $N$ times those for a single oscillator. 
The $q$ dependence of the energy and the that of heat capacity are quite weak.
It seems that finding the difference is difficult.
In contrast, the Tsallis entropy divided by $N$ decreases with $N$ and the Tsallis entropy decreases with $q$.
The differences can be seen at high equilibrium temperature. 

It is not trivial whether the argument of the $q$-exponential, $1+(q-1) \betaPh (E_n - \Uthree)$, is positive in the Tsallis-3 statistics of $q>1$:
the condition of the positivity is not trivial because of the existence of $\Uthree$.
This condition is rewritten as $1-(q-1) \betaPh \URthree > 0$. 
We found the inequality $(\URthree/\aeq)/N < \tPh$ from numerical calculations for $a_1 = \cdots = a_N$.
This inequality is rewritten as $\betaPh \URthree < N$.
With this result, we have the inequality $1-(q-1) \betaPh \URthree > 1 - (q-1) N$.
Therefore, the condition of the positivity is satisfied when $(q-1)N$ is less than one.
The condition, $(q-1)N<1$, already appeared in the numerical calculations to choose the value of $q$.

The entropic parameter $q$ is limited in the Tsallis statistics.
The limitation is $N/(N+1) < q < 1$ in the Tsallis-1 statistics. 
The limitation is $1< q < (N+1)/N $ in the Tsallis-2 statistics and the Tsallis-3 statistics.    
The requirements that the distributions are power-like lead to the limitations of $q$:
$q$ is less than one in the Tsallis-1 statistics, and $q$ is greater than one in the Tsallis-2 statistics and the Tsallis-3 statistics. 
The relations between $q$ and $N$ arise from the requirements for the Barnes zeta function. 
The parameter $q$ approaches one as $N$ goes to infinity.

The results indicate that the Tsallis-1 statistics and the Tsallis-3 statistics are preferable for systems that are invariant to energy shifts.
The well-known relation between the energy $U$ and the average energy level $\overline{n}$ 
for the independent oscillators is hold in the Tsallis-1 statistics and the Tsallis-3 statistics,
while the relation is modified in the Tsallis-2 statistics.
As explicitly shown in the numerical calculations, the physical quantities depend on the zero-point energy in the Tsallis-2 statistics.
In the numerical calculations where the zero-point energies are nonzero,
the entropies $\STtwo$ do not approach zero as the scaled temperature $\scTtwo$ goes to zero. 
In addition, the expectation values of the unit operator $\hat{1}$ are one in the Tsallis-1 statistics and the Tsallis-3 statistics.
These facts indicate that the Tsallis-1 statistics and the Tsallis-3 statistics are advantageous for describing systems that are invariant to energy shifts.
It is noted again that the Tsallis-1 statistics applied in this paper is different from that applied in the earlier work \cite{Tsallis:1988}.

In this paper, we studied the system of $N$-independent oscillators in the Tsallis statistics
in the case that the distribution is power-like, 
using the expansion of the Barnes zeta function with the Hurwitz zeta function.
Our results can be applied to the system whose energy is given by $\displaystyle\sum_j (a_j n_j + b_j)$.
I hope that this work is helpful in the future studies related to unconventional statistics.

\bigskip

\noindent
\textbf{Funding}\quad
This research received no specific grant from any funding agency in the public, commercial, or not-for-profit sectors.

\medskip

\noindent
\textbf{Data availability}\quad
This manuscript has no associated data or the data will not be deposited.
[Authors' comment: This study is theoretical, and the graphs were drawn with the equations given in this paper.].

\medskip

\noindent
\textbf{Conflict of interest}\quad
The author declares no competing interest.

\appendix
\section{The expansion of the Barnes zeta function with the Hurwitz zeta function}
\label{sec:expansion}
The Hurwitz zeta function $\zeta_H(s,a)$ and the Barnes zeta function $\zeta_B(s,a|a_1, \cdots, a_N)$ are defined by
\begin{subequations}
\begin{align}
& \zeta_H(s,a) = \sum_{n=0}^{\infty} \frac{1}{(n+a)^s}, \\
& \zeta_B(s,a|a_1, \cdots, a_N) = \sum_{n_1=0}^{\infty} \sum_{n_2=0}^{\infty}\cdots \sum_{n_N=0}^{\infty}
  \frac{1}{(a + a_1 n_1 + a_2 n_2 + \cdots + a_N n_N)^s}.
  \label{Appendix:def:BarnesZeta}
\end{align}
\end{subequations}
The Barnes zeta function $\zeta_B(s,a|a_1, \cdots, a_N)$ can be expanded with the Hurwitz zeta function $\zeta_H(s,a)$ \cite{Elizalde1989, Oprisan}.
The following equations are described in the reference \cite{Oprisan}:
\begin{subequations}
\begin{align}
  \zeta_B(s, a| a_1, \cdots, a_N) 
  &= \frac{1}{\Gamma(s)} \sum_{n_1, \cdots, n_N =0}^{\infty} \int_0^{\infty} dt\ t^{s-1} e^{-t (a+a_1 n_1 + \cdots + a_N n_N)} \label{zeta-eq:1} \\
  &= \frac{1}{\Gamma(s)}  \int_0^{\infty} dt\ t^{s-1} e^{-at}
  \left( \frac{1}{(1-e^{-a_N t})} + \sum_{k=1}^{N-1} e^{-a_k t} \prod_{j=k}^{N} (1-e^{-a_j t})^{-1} \right) \label{zeta-eq:2}\\
  &= \frac{1}{(a_N)^s} \left[
  \zeta_H(s, d) + \sum_{p=1}^{\infty} \sum_{1 \le k_1 \le \cdots \le k_p \le N-1} \zeta_H(s, d+d_{k_1}+\cdots+d_{k_p})
  \right]
  \label{zeta-eq:3}\\
  &\equiv \frac{1}{(a_N)^s} \sum_{p=0}^{\infty} \sum_{1 \le k_1 \le \cdots \le k_p \le N-1} \zeta_H(s, d+d_{k_1}+\cdots+d_{k_p}) \label{zeta-eq:4}, 
\end{align}
\end{subequations}
where $d_j=a_j/a_N$ and  $d=a/a_N$ $(j=1,2,\cdots,N)$. In these equations, the following notation is adopted:
\begin{align}
  \sum_{1 \le k_1 \le \cdots \le k_p \le N-1} \equiv \sum_{k_1 = 1}^{N-1} \sum_{k_2 = k_1}^{N-1} \cdots \sum_{k_p = k_{p-1}}^{N-1} . 
\end{align}

The equation \eqref{zeta-eq:1} is easily derived by applying the following formula \cite{Abramowitz,Gradshteyn} to the Barnes zeta function:
\begin{align} 
\frac{1}{x^s} = \frac{1}{\Gamma(s)} \int_0^{\infty} dt t^{s-1} e^{-xt} ,  
\label{formula:1}
\end{align}
where $\Gamma(s)$ is the Gamma function.

In the present appendix, we provide the brief derivation according to the strategy in the reference \cite{Elizalde1989}
and we give a different method for deriving the expansion.

\subsection{Derivation according to the strategy given in the previous work}
In this subsection, we derive Eqs.~\eqref{zeta-eq:2} and \eqref{zeta-eq:3} briefly
according to the strategy described in the reference \cite{Elizalde1989}.

Equation~\eqref{zeta-eq:1} can be rewritten with the following equation:
\begin{align}
  \sum_{n_1=0}^{\infty} \sum_{n_2=0}^{\infty} \cdots \sum_{n_N=0}^{\infty} e^{-ta_1 n_1-ta_2 n_2 \cdots -ta_N n_N}
  = \frac{1}{ (1-e^{-ta_N})(1-e^{-ta_{N-1}}) \cdots (1-e^{-ta_1})} .
  \label{eqn:simple_sum}
\end{align}
In addition, we use the following equation:
\begin{align}
  \prod_{j=l}^{N} (1-e^{-a_j t})^{-1} = \prod_{j=l+1}^{N} (1-e^{-a_j t})^{-1} + e^{-a_l t}  \prod_{j=l}^{N} (1-e^{-a_j t})^{-1} . 
\end{align}
This equation leads to
\begin{align}
  \prod_{j=l}^{N} (1-e^{-a_j t})^{-1}  = (1-e^{-a_N t})^{-1} + \sum_{k=l}^{N-1} e^{-a_k t}  \prod_{j=k}^{N} (1-e^{-a_j t})^{-1} . 
  \label{el:2}
\end{align}
Using Eqs.~\eqref{eqn:simple_sum} and \eqref{el:2} with $l=1$, we have Eq.~\eqref{zeta-eq:2}.

We show the brief derivation of Eq.~\eqref{zeta-eq:3}. 
We define functions $J_0^{k_0}$, $J_l^{k_0}$ ($l \ge 1$), and $\tilde{R}_{p}^{k_0}$:
\begin{subequations}
\begin{align}
  J_0^{k_0} &= \frac{1}{\Gamma(s)} \int_0^{\infty}\ dt\ t^{s-1} e^{-at} (1-e^{-a_N t})^{-1},\\
  J_l^{k_0} &= \frac{1}{\Gamma(s)} \int_0^{\infty}\ dt\ t^{s-1} e^{-at} (1-e^{-a_N t})^{-1}
  \sum_{k_1=k_0}^{N-1} \sum_{k_2=k_1}^{N-1} \cdots \sum_{k_{l}=k_{l-1}}^{N-1} e^{-a_{k_1} t} e^{-a_{k_2} t} \cdots e^{-a_{k_{l}} t} \quad (l \ge 1), \\
  \tilde{R}_{p}^{k_0} &=
  \frac{1}{\Gamma(s)} \int_0^{\infty}\ dt\ t^{s-1} e^{-at} 
  \sum_{k_1=k_0}^{N-1} \sum_{k_2=k_1}^{N-1} \cdots \sum_{k_p=k_{p-1}}^{N-1}  e^{-a_{k_1} t} e^{-a_{k_2} t} \cdots e^{-a_{k_p} t} \prod_{j=k_p}^{N} (1-e^{-a_j t})^{-1} . 
\end{align}
\end{subequations}
We note that the superscript $k_0$ is attached to $J_0$ to use the same notation as $J_l^{k_0}$.
With these functions, the Barnes zeta function is represented by using Eq.~\eqref{el:2} recursively:
\begin{align}
  \zeta_B(s, a| a_1, \cdots, a_N) =  \sum_{j=0}^{p-1} J_j^{k_0=1} + \tilde{R}_p^{k_0=1} . 
  \label{eqn:Barnes_zeta_with_J_R}
\end{align}
Expanding $(1-e^{-a_N t})^{-1}$, we easily calculate the functions, $J_0^{k_0}$ and $J_l^{k_0}$, as
\begin{subequations}
\begin{align}
  &J_0^{k_0} = \frac{1}{(a_N)^s} \zeta_H(s,d), \label{eq:J_0}\\
  &J_l^{k_0} = \sum_{k_1=k_0}^{N-1} \sum_{k_2=k_1}^{N-1} \cdots \sum_{k_l=k_{l-1}}^{N-1}  \frac{1}{(a_N)^s} \zeta_H(s,d+d_{k_1}+\cdots+d_{k_l}), \label{eq:J_l}
\end{align}
\end{subequations}
where $d=a/a_N$ and $d_j = a_j/a_N$. 
Therefore, we obtain Eq.~\eqref{zeta-eq:3} after showing the vanishment of $\tilde{R}_p^{k_0=1}$ in the limit $p \rightarrow \infty$. 
The function $\tilde{R}_p^{k_0=1}$ is rewritten:
\begin{subequations}
\begin{align}
& \tilde{R}_p^{k_0=1} = (a_N)^{-s}  R_p^{k_0=1}, \label{tildeR:R}\\
& R_p^{k_0=1} = 
\frac{1}{\Gamma(s)} \sum_{1 \le k_1 \le \cdots \le k_p \le {N-1}} 
\int_0^{\infty}\ dt\ t^{s-1}  e^{-(d + d_{k_1} + d_{k_2} + \cdots + d_{k_p}) t} \prod_{j=k_p}^{N} (1-e^{-d_j t})^{-1} . 
\end{align}
\end{subequations}
Therefore, we show that $R_p^{k_0=1}$ approaches zero as $p$ goes to infinity. 

Without loss of generality, we assume that $d_j$ is equal to or greater than one, because $a_j$ can be arranged in decreasing order
in the Barnes zeta function: $a_1 \ge a_2 \ge \cdots \ge a_N$. 
We note that $d$ is greater than zero, because $a$ in Eq.~\eqref{Appendix:def:BarnesZeta} is greater than zero.
As suggested in the reference,
the integral $R_p^{k_0=1}$ is divided into two integrals $R_{\mathrm{L},p}^{k_0=1}$ and $R_{\mathrm{H},p}^{k_0=1}$.
The region of $R_{\mathrm{L},p}^{k_0=1}$ is $0 \le t \le \varepsilon$ and 
the region of $R_{\mathrm{H},p}^{k_0=1}$ is $\varepsilon \le t < \infty $, where $\varepsilon$ satisfies $0< \varepsilon <1$.
We have 
\begin{subequations}
\begin{align}
  \left| R_p^{k_0=1} \right| &= \left| R_{\mathrm{L},p}^{k_0=1} + R_{\mathrm{H},p}^{k_0=1} \right|
  \le \left| R_{\mathrm{L},p}^{k_0=1} \right| + \left| R_{\mathrm{H},p}^{k_0=1} \right| ,\\
  \left| R_{\mathrm{L},p}^{k_0=1} \right| &\le 
  \left| \frac{1}{\Gamma(s)}
  \sum_{1\le k_1 \le \cdots \le k_p \le N-1} 
  \int_0^{\varepsilon}\ dt\ t^{s-1}  e^{-td} \frac{1}{(1-e^{-t})^N} e^{-(d_{k_1} + d_{k_2} + \cdots + d_{k_p}) t}
  \right| ,\\
  \left| R_{\mathrm{H},p}^{k_0=1} \right| &\le 
  \left| \frac{1}{\Gamma(s)}
  \sum_{1\le k_1 \le \cdots \le k_p \le N-1}
  \int_{\varepsilon}^{\infty}\ dt\ t^{s-1}  e^{-td} \frac{1}{(1-e^{-t})^N} e^{-(d_{k_1} + d_{k_2} + \cdots + d_{k_p}) t}
  \right|, 
\end{align}
\end{subequations}
where $d>0$ and $d_j \ge 1$. 

First, we deal with $R_{\mathrm{L},p}^{k_0=1}$.
The following equation is easily derived:
\begin{align}
\sum_{1 \le k_1 \le \cdots \le k_p \le N} 1 = \left( \begin{array}{c} N-1 + p \\ p \end{array} \right). 
\label{eqn:combination}
\end{align}
Using Eq.~\eqref{eqn:combination} and expanding the integral interval,
we obtain 
\begin{align}
\left| R_{\mathrm{L},p}^{k_0=1} \right| \le 
\left| 
\frac{1}{\Gamma(s)}  \frac{1}{(1 - \varepsilon/2)^N} \frac{1}{(N-2)!} 
\frac{(N+p-2)!}{p!} 
\int_0^{\infty}\ dt\ t^{s-N-1} e^{-pt} \right| . 
\end{align}
Applying the Stirling's formula, for $s>N$, we have 
\begin{align}
\left| R_{\mathrm{L},p}^{k_0=1} \right| \le 
\frac{\Gamma(s-N)}{\Gamma(s)}  \frac{1}{(1 - \varepsilon/2)^N (N-2)!}
\ p^{2N-2-s} (1 + O(p^{-1})) . 
\label{R_L:eval}
\end{align}
The right-hand side of Eq.~\eqref{R_L:eval} goes to zero as $p$ goes to infinity for $s>2N-2$ and $s>N$.

Next, we estimate the remaining part $R_{\mathrm{H},p}^{k_0=1}$.
For $s>0$, we have
\begin{align}
  \left| R_{\mathrm{H},p}^{k_0=1} \right|
  &\le \frac{1}{(1-e^{-\varepsilon})^N (N-2)!}\ p^{N-s-2} (1+O(p^{-1})) . 
\label{R_H:eval}
\end{align}
The right-hand side of Eq.~\eqref{R_H:eval} goes to zero as $p$ goes to infinity for $s>N-2$ and $s>0$.

We obtain the expansion of the Barnes zeta function, Eq.~\eqref{zeta-eq:3},
from Eqs.~\eqref{eqn:Barnes_zeta_with_J_R}, \eqref{eq:J_0}, and \eqref{eq:J_l}, 
because $\tilde{R}_p^{k_0=1}$ converges to zero for $s>2N-2$ with $N \ge 2$.
Equation~\eqref{zeta-eq:3} is trivial for $N=1$.
The convergence of $\tilde{R}_p^{k_0=1}$ is easily found from Eq.~\eqref{tildeR:R}, 
because $R_p^{k_0=1}$ converges to zero as $p$ goes to infinity from Eqs.~\eqref{R_L:eval} and ~\eqref{R_H:eval} .

\subsection{A different method for deriving the equations}
We give a different method for deriving Eqs.~\eqref{zeta-eq:2} and \eqref{zeta-eq:3} in this subsection, 
because this derivation might be helpful in other studies.

\subsubsection{Derivation of Eq.~\eqref{zeta-eq:2}}
We derive Eq.~\eqref{zeta-eq:2} by showing the following equation.
\begin{align}
  \sum_{n_1=0,n_2=0,\cdots,n_N=0}^{\infty} e^{-t(a_1 n_1 + \cdots + a_N n_N)}
  = \frac{1}{(1-e^{-a_1t}) \cdots (1-e^{-a_Nt}) }
  = L_N , 
  \label{My:2:eq_to_proof}
\end{align}
where $L_N$ is defined by
\begin{align}
  L_N = (1-e^{-a_N t})^{-1} + \sum_{k=1}^{N-1} e^{-a_k t} \prod_{j=k}^{N} (1-e^{-a_j t})^{-1} .
  \label{expression_J_N}
\end{align}
We use the following conventions for empty sum and empty product:
\begin{subequations}
\begin{align}
  &\sum_{k=1}^0 A_k = 0, \\
  &\prod_{k=1}^0 A_k = 1.
\end{align}
\end{subequations}
With these conventions, we use the expression of $L_N$, Eq.~\eqref{expression_J_N}, for $N \ge 1$.

Equation~\eqref{My:2:eq_to_proof} is correct for $N=1$ apparently.
To proceed with the calculation, we define the function $K_{N-1}$ by
\begin{subequations}
\begin{align}
  &L_N = \frac{1}{(1-e^{-a_1 t}) \cdots (1-e^{-a_N t}) } K_{N-1}, \\
  &K_{N-1} = (1-e^{-a_1 t}) \cdots (1-e^{-a_{N-1} t}) + \sum_{k=1}^{N-1} e^{-a_k t} \prod_{j=1}^{k-1} (1-e^{-a_j t}). 
\end{align}
\end{subequations}
We prove Eq.~\eqref{My:2:eq_to_proof} by showing $K_{N-1} = 1$ for $N \ge 2$.  

It is easily shown by calculating $K_{N-1}$ directly that $K_{N-1}$ equals one for $N=2$ and $N=3$.
Therefore, we attempt to show that $K_{N}$ equals one under the assumption that $K_{N-1}$ equals one. 
The function $K_{N}$ is calculated as follows:
\begin{align}
  K_N &= (1-e^{-a_1 t}) \cdots (1-e^{-a_N t}) + \sum_{k=1}^{N} e^{-a_k t} \prod_{j=1}^{k-1} (1-e^{-a_j t}) \nonumber \\
  &= (1-e^{-a_1 t}) \cdots (1-e^{-a_N t})
  + e^{-a_N t} \prod_{j=1}^{N-1} (1-e^{-a_j t})
  + \sum_{k=1}^{N-1} e^{-a_k t} \prod_{j=1}^{k-1} (1-e^{-a_j t}). 
\end{align}
Using the assumption $K_{N-1} = 1$, we have 
\begin{align}
  K_N  &= (1-e^{-a_1 t}) \cdots (1-e^{-a_N t})
  + e^{-a_N t} \prod_{j=1}^{N-1} (1-e^{-a_j t})
  + \left[ 1 - (1-e^{-a_1 t}) \cdots (1-e^{-a_{N-1} t}) \right] = 1.
\end{align}
By mathematical induction, we conclude that $K_{N-1}$ equals one for $N \ge 2$. 
From these calculations, Eq.~\eqref{My:2:eq_to_proof} is proven, and therefore Eq.~\eqref{zeta-eq:2} is derived. 

\subsubsection{Derivation of Eq.~\eqref{zeta-eq:3}}
We attempt to show the following equation:
\begin{align}
  & \frac{1}{\Gamma(s)} \int_0^{\infty} dt t^{s-1} e^{-at} \Bigg[ (1-e^{-a_N t})^{-1} + \sum_{k=1}^{N-1} e^{-a_k t} \prod_{j=k}^{N} (1-e^{-a_j t} )^{-1} \Bigg] \nonumber \\
  & = (a_N)^{-s} \Bigg[ \zeta(s,d) + \sum_{p=1}^{\infty} \sum_{1 \le k_1 \le \cdots \le k_p \le N-1} \zeta_H(s,d+d_{k_1}+\cdots+d_{k_p}) \Bigg] .
  \label{eqn:my:3}
\end{align}
We rewrite Eq.~\eqref{eqn:my:3} for the derivation.
By changing of the variable $u=a_N t$, the left-hand side of Eq.~\eqref{eqn:my:3} is given as follows:
\begin{align}
  &\frac{1}{\Gamma(s)} \int_0^{\infty} dt\ t^{s-1} e^{-at} \Big[ (1-e^{-a_N t})^{-1} + \sum_{k=1}^{N-1} e^{-a_k t} \prod_{j=k}^{N} (1-e^{-a_j t} )^{-1} \Big] \nonumber \\
  &= \frac{1}{(a_{N})^{s}} \frac{1}{\Gamma(s)} \int_0^{\infty} du\  u^{s-1} e^{-ud} 
  \Bigg[ (1-e^{-u})^{-1} + \sum_{k=1}^{N-1} e^{-ud_k} \prod_{j=k}^{N} (1-e^{-u d_j})^{-1} \Bigg].
\end{align}
By applying Eq.~\eqref{formula:1}, the right-hand side of Eq.~\eqref{eqn:my:3} is given as follows:
\begin{align}
  & \frac{1}{(a_N)^{s}} \Bigg[ \zeta(s,d) + \sum_{p=1}^{\infty} \sum_{1 \le k_1 \le \cdots \le k_p \le N-1} \zeta_H(s,d+d_{k_1}+\cdots+d_{k_p}) \Bigg] \nonumber \\
  &=  \frac{1}{(a_N)^{s}} \frac{1}{\Gamma(s)} \int_0^{\infty} dt\ t^{s-1} (1-e^{-t})^{-1}
  \Bigg[e^{-d t} + \sum_{p=1}^{\infty} \sum_{1\le k_1 \le \cdots \le k_{p} \le N-1} e^{-(d+d_{k_1}+\cdots+d_{k_p}) t} \Bigg] .
\end{align}
Equation~\eqref{eqn:my:3} is rewritten:
\begin{align}
&\int_0^{\infty} du\  u^{s-1} e^{-ud} \Bigg[ (1-e^{-u})^{-1} + \sum_{k=1}^{N-1} e^{-ud_k} \prod_{j=k}^{N} (1-e^{-u d_j})^{-1} \Bigg] \nonumber \\
&= \int_0^{\infty} dt\ t^{s-1} (1-e^{-t})^{-1} \Bigg[e^{-d t} + \sum_{p=1}^{\infty} \sum_{1\le k_1 \le \cdots \le k_{p} \le N-1} e^{-(d+d_{k_1}+\cdots+d_{k_p}) t} \Bigg].
\label{eqn:relation_integrals}
\end{align}
Equation~\eqref{eqn:relation_integrals} for $N=1$ is trivial. 
Therefore, for $N \ge 2$, we attempt to prove the equation:
\begin{align}
\sum_{k=1}^{N-1} e^{-d_k t} \prod_{j=k}^{N} (1-e^{-d_j t})^{-1} = (1-e^{-t})^{-1}  \sum_{p=1}^{\infty} \sum_{1\le k_1 \le \cdots \le k_{p} \le N-1} e^{-(d_{k_1}+\cdots+d_{k_p}) t} .
\end{align}
All the terms contain $(1-e^{-t})$ in the left-hand side of the above equation because of $d_N = 1$.
Therefore, we should prove the following equation: 
\begin{align}
\sum_{k=1}^{N-1} e^{-d_k t} \prod_{j=k}^{N-1} (1-e^{-d_j t})^{-1} = \sum_{p=1}^{\infty} \sum_{1\le k_1 \le \cdots \le k_{p} \le N-1} e^{-(d_{k_1}+\cdots+d_{k_p}) t} .
\label{eqn:my:3:mod}
\end{align}
Equation~\eqref{eqn:my:3} is proven if Eq.~\eqref{eqn:my:3:mod} is proven.

We attempt to prove the following equation with $b_j > 0$,
where we use the parameters $b_j$ to avoid confusion: 
\begin{align}
\sum_{k=1}^{N-1} e^{-b_k t} \prod_{j=k}^{N-1} (1-e^{-b_j t})^{-1} = \sum_{p=1}^{\infty} \sum_{1\le k_1 \le \cdots \le k_{p} \le N-1} e^{-(b_{k_1}+\cdots+b_{k_p}) t}.
\label{eqn:my:4}
\end{align}
To simplify the equation, we define the functions $f(k_1,\cdots,k_p)$ and $g(p,N-1)$ by
\begin{subequations}
\begin{align}
 & f(k_1,\cdots,k_p) = \exp(-(b_{k_1}+b_{k_2}+\cdots+b_{k_p}) t), \label{eqn:simplify:f}\\
 & g(p,N-1) = \sum_{k_1=1}^{N-1} \sum_{k_2=k_1}^{N-1} \cdots \sum_{k_p=k_{p-1}}^{N-1}  f(k_1, \cdots, k_p). \label{eqn:simplify:g}
\end{align}
\end{subequations}
Equation~\eqref{eqn:my:4} is rewritten with Eqs.~\eqref{eqn:simplify:f} and \eqref{eqn:simplify:g} as
\begin{align}
\sum_{k=1}^{N-1} e^{-b_k t} \prod_{j=k}^{N-1} (1-e^{-b_j t})^{-1} = \sum_{p=1}^{\infty} g(p, N-1) .
\label{eqn:my:5}
\end{align}
At this moment, our problem has been transformed into showing Eq.~\eqref{eqn:my:5}. 

We attempt to derive Eq.~\eqref{eqn:my:5} by calculating the right-hand side of Eq.~\eqref{eqn:my:5}.
We define $X^{(N-1)}$ by
\begin{align}
  X^{(N-1)} = \sum_{p=1}^{\infty} g(p, N-1) .
\end{align}
We give the recurrence relation of $X^{(N-1)}$ to derive Eq.~\eqref{eqn:my:5}.
The function $g(p,N-1)$ has the following relation which is obtained by dealing with the sum of $k_p$:
\begin{align}
  g(p,N-1) &=
  \sum_{1\le k_1 \le \cdots \le k_{p-1} \le 1} f(k_1, k_2, \cdots, k_{p-1}, k_p=1)
  + \sum_{1\le k_1 \le \cdots \le k_{p-1} \le 2} f(k_1, k_2, \cdots, k_{p-1}, k_p=2)
  \nonumber \\ & \qquad
  + \cdots + \sum_{1\le k_1 \le \cdots \le k_{p-1} \le N-1} f(k_1, k_2, \cdots, k_{p-1}, k_p=N-1)
  \nonumber \\ &
  = g(p-1,1) e^{-b_1 t} + g(p-1, 2) e^{-b_2 t} + \cdots + g(p-1, N-1) e^{-b_{N-1} t}.
\label{eqn:expansion_of_g}
\end{align}
From Eq.~\eqref{eqn:expansion_of_g}, we have
\begin{align}
  \sum_{p=1}^{\infty} g(p+1,N-1) = X^{(1)} e^{-b_1 t} + X^{(2)} e^{-b_2 t} + \cdots + X^{(N-1)} e^{-b_{N-1} t}.
  \label{eqn:gX_relation}
\end{align}
The left-hand side of Eq.~\eqref{eqn:gX_relation} is given by $X^{(N-1)} - g(1,N-1)$.
Therefore, we obtain
\begin{align}
  (1 - e^{-b_{N-1} t}) X^{(N-1)} - e^{-b_{N-2} t} X^{(N-2)} - \cdots - e^{-b_2 t} X^{(2)} - e^{-b_1 t} X^{(1)} = g(1,N-1).  
\label{eqn:recurrence_rel}
\end{align}
For $N=2$, we find the relation easily: $(1 - e^{-b_{1} t}) X^{(1)} = g(1,1)$. 
We obtain the following equation with the matrix $M^{(N-1)}$ 
from Eq.~\eqref{eqn:recurrence_rel}:
\begin{subequations}
\begin{align}
  &
  M^{(N-1)} \left( \begin{array}{c} X^{(N-1)} \\ X^{(N-2)} \\ \vdots \\ X^{(1)} \end{array} \right)
  = 
  \left( \begin{array}{c} g(1,N-1) \\ g(1,N-2) \\ \vdots \\\ g(1,1) \end{array} \right),\\
  &
  M^{(N-1)} =
  \left(
  \begin{array}{ccccccc}
    (1 - e^{-b_{N-1} t}) & -e^{-b_{N-2} t} & -e^{-b_{N-3} t} &  & \cdots &  & -e^{-b_{1} t} \\
    0 & (1 - e^{-b_{N-2} t}) & -e^{-b_{N-3} t} &   & \cdots &  & -e^{-b_{1} t} \\
    0 & 0 & (1 - e^{-b_{N-3} t}) &  & & & \vdots \\
    \vdots & \vdots & 0 &  & & & \vdots \\    
    \vdots & \vdots & \vdots & &  & \ddots &  -e^{-b_{1} t} \\
    0 & 0 & 0 & \cdots & \cdots & 0 & (1 - e^{-b_{1} t})
  \end{array}
  \right).
\end{align}
We define the matrix $K^{(N-1)}$ by 
\begin{align}
  & K^{(N-1)} =
  \left(
  \begin{array}{ccccccc}
    g(1,N-1) & -e^{-b_{N-2} t} & -e^{-b_{N-3} t} &  &\cdots & & -e^{-b_{1} t} \\
    g(1,N-2) & (1 - e^{-b_{N-2} t}) & -e^{-b_{N-3} t} & & \cdots & & -e^{-b_{1} t} \\
    g(1,N-3) & 0 & (1 - e^{-b_{N-3} t}) &  & & & \vdots \\
    \vdots & \vdots & 0 &  & & & \vdots \\    
    \vdots & \vdots & \vdots & & & \ddots &  -e^{-b_{1} t} \\
    g(1,1) & 0 & 0 & \cdots & & 0 & (1 - e^{-b_{1} t})
  \end{array}
  \right).
\end{align}
\end{subequations}
The quantity $X^{(N-1)}$ is represented by
\begin{align}
  X^{(N-1)} = \frac{|K^{(N-1)}|}{|M^{(N-1)}|}, 
\end{align}
where $|M|$ represents the determinant of the matrix $M$.
The determinant of $M^{(N-1)}$ is given by
\begin{align}
  |M^{(N-1)}| = \prod_{j=1}^{N-1} (1-e^{-b_j t}) . 
\end{align}

The determinant of $K^{(N-1)}$ is calculated with ${\displaystyle g(1,N-1) = \sum_{k_1=1}^{N-1} e^{-b_{k_1} t}}$:
\begin{align}
  |K^{(N-1)}| &=
  \left|
  \begin{array}{cccccc}
    e^{-b_{N-1} t }  & -e^{-b_{N-2} t} & -e^{-b_{N-3} t} &\cdots &\cdots & -e^{-b_{1} t} \\
    1 & (1 - e^{-b_{N-2} t}) & -e^{-b_{N-3} t} &\cdots & \cdots & -e^{-b_{1} t} \\
    1 & 0 & (1 - e^{-b_{N-3} t}) &  & & \vdots \\
    \vdots & \vdots & 0 &  & & \vdots \\    
    \vdots & \vdots & \vdots &\ddots & \ddots &  -e^{-b_{1} t} \\
    1 & 0 & 0 & \cdots  & 0 & (1 - e^{-b_{1} t})
  \end{array}
  \right| 
  \nonumber \\
  &= 
  \left|
  \begin{array}{cccccc}
    (e^{-b_{N-1} t }-1) & -1 & 0 & 0 &\cdots & 0\\
    0 & (1 - e^{-b_{N-2} t}) & -1 & 0 & \cdots & 0\\
    0 & 0 & (1 - e^{-b_{N-3} t}) & -1 & & \vdots \\
    \vdots & \vdots & 0 &      &\ddots & \vdots \\    
    0 & \vdots & \vdots & \ddots & \ddots &  -1 \\
    1 & 0 & 0 & \cdots &  0 & (1 - e^{-b_{1} t})
  \end{array}
  \right| . 
\end{align}
Here we define $L^{(N-2)}$ and $H^{(N-3)}$ by 
\begin{align}
&L^{(N-2)} = \left(
  \begin{array}{ccccc}
     (1 - e^{-b_{N-2} t}) & -1 & 0 & \cdots & 0\\
     0 & (1 - e^{-b_{N-3} t}) & -1 & & \vdots \\
     \vdots & 0 &  & & \vdots \\    
     \vdots & \vdots & & \ddots &  -1 \\
     0 & \cdots & \cdots & 0 & (1 - e^{-b_{1} t})
  \end{array}
  \right),
  \\
&H^{(N-3)} = \left(
  \begin{array}{cccccc}
     0 & -1 &  0 & \cdots & \cdots & 0\\
     0 & (1 - e^{-b_{N-3} t}) &  -1 & & & \vdots \\
     \vdots & 0      & (1 - e^{-b_{N-4} t}) & & & \vdots \\    
            &        &  & \ddots &                  &  0 \\
     \vdots & \vdots & & 0& (1 - e^{-b_{2} t}) &  -1 \\
     1 & 0  & \cdots & 0 & 0& (1 - e^{-b_{1} t})
  \end{array}
  \right).
\end{align}
With these matrices, the determinant of $K^{(N-1)}$ is given by
\begin{align}
  |K^{(N-1)}| = (e^{-b_{N-1} t} - 1) |L^{(N-2)}| + |H^{(N-3)}| .
\end{align}
It is easily found that
\begin{subequations}
\begin{align}
  &|L^{(N-2)}| = \prod_{j=1}^{N-2} (1-e^{-b_j t}), \\
  &|H^{(N-3)}| = |H^{(N-4)}| = \cdots = |H^{(2)}|
  = \left|\begin{array}{ccc}
  0 & -1 & 0 \\
  0 & 1-e^{-b_2 t} & -1 \\
  1 & 0 & 1-e^{-b_1 t}
  \end{array}\right| = 1 .
\end{align}
\end{subequations}
Therefore, we have
\begin{align}
|K^{(N-1)}| = 1 - \prod_{j=1}^{N-1} (1-e^{-b_j t}) . 
\end{align}
As a result, we obtain
\begin{align}
X^{(N-1)} = \frac{1}{\displaystyle\prod_{j=1}^{N-1} (1-e^{-b_j t})} - 1 .
\end{align}
This equation leads to the recurrence relation:
\begin{align}
X^{(N)} = \frac{1}{(1-e^{-b_{N} t})} X^{(N-1)} + \frac{e^{-b_{N} t}}{(1-e^{-b_{N} t})} .
\label{eqn:XnXn-1}
\end{align}

By mathematical induction, we attempt to prove Eq.~\eqref{eqn:my:5} using Eq.~\eqref{eqn:XnXn-1}.
Equation~\eqref{eqn:my:5} is rewritten as 
\begin{align}
\sum_{k=1}^{N-1} e^{-b_k t} \prod_{j=k}^{N-1} (1-e^{-b_j t})^{-1} = X^{(N-1)}.
\label{eqn:my:5:new}
\end{align}
Equation~\eqref{eqn:my:5:new} for $N=2$ is easily demonstrated.
Therefore, under the assumption that Eq.~\eqref{eqn:my:5:new} is correct, we attempt to prove the following equation:
\begin{align}
\sum_{k=1}^{N} e^{-b_k t} \prod_{j=k}^{N} (1-e^{-b_j t})^{-1} = X^{(N)}.
\end{align}
The quantity $X^{(N)}$ is calculated with Eq.~\eqref{eqn:XnXn-1}.
\begin{align}
  X^{(N)} &= \frac{1}{(1-e^{-b_{N} t})} X^{(N-1)} + \frac{e^{-b_{N} t}}{(1-e^{-b_{N} t})} 
  =  \frac{1}{(1-e^{-b_{N} t})} \left[\sum_{k=1}^{N-1} e^{-b_k t} \prod_{j=k}^{N-1} (1-e^{-b_j t})^{-1}\right] + \frac{e^{-b_{N} t}}{(1-e^{-b_{N} t})} \nonumber \\
  &=   \left[\sum_{k=1}^{N-1} e^{-b_k t} \prod_{j=k}^{N} (1-e^{-b_j t})^{-1}\right] + \frac{e^{-b_{N} t}}{(1-e^{-b_{N} t})}
  =   \sum_{k=1}^{N} e^{-b_k t} \prod_{j=k}^{N} (1-e^{-b_j t})^{-1} .
\end{align}
Therefore, Eq.~\eqref{eqn:my:5:new} is correct for $N \ge 2$. 

Equation~\eqref{eqn:my:4} is proven by showing Eq.~\eqref{eqn:my:5:new} for $N \ge 2$.
As a result, Eq.~\eqref{eqn:my:3:mod} is proven for $N \ge 2$.
Therefore, Eq.~\eqref{eqn:my:3} for $N \ge 1$ is derived.
That is, Eq.~\eqref{zeta-eq:3} is proven.

\subsection{The case of $a_1 = a_2 = \cdots = a_N$}
\label{subsec:barnes:equal}

We treat the case where $a_1$, $a_2$, $\cdots$, $a_N$ are all equal in this subsection.  

By setting $a_1 = a_2 = \cdots = a_N$ in Eq.~\eqref{zeta-eq:3}, we have
\begin{align}
  \left. \zeta_B(s, a| a_1, \cdots, a_N) \right|_{a_1=a_2=\cdots=a_N} 
  &= \frac{1}{(a_1)^s} \left[\zeta_H(s, d) + \sum_{p=1}^{\infty} \zeta_H(s, d + p ) \sum_{1 \le k_1 \le \cdots \le k_p \le N-1} 1 \right], 
\end{align}
because of $d=a/a_N=a/a_1$ and $d_j = a_j/a_N = 1$.
With Eq.~\eqref{eqn:combination}, we obtain
\begin{align}
  \left. \zeta_B(s, a| a_1, \cdots, a_N) \right|_{a_1=a_2=\cdots=a_N} 
  &= \frac{1}{(a_1)^s} \left[\zeta_H(s, d) + \sum_{p=1}^{\infty} \left( \begin{array}{c} N + p -2 \\ p \end{array} \right) \zeta_H(s, d + p )  \right]. 
\end{align}
This equation was already given in the previous papers \cite{Elizalde1989, Oprisan}.


\end{document}